\documentclass[acmtog,nonacm]{acmart}
\renewcommand\footnotetextcopyrightpermission[1]{} 

\usepackage{booktabs} 

\usepackage{multirow}
\usepackage{array}
\usepackage[ruled]{algorithm2e} 
\usepackage{graphicx}
\usepackage{xcolor}
\usepackage{amsmath}
\usepackage{hyperref}  

\usepackage{cleveref}

\SetAlFnt{\small}
\SetAlCapFnt{\small}
\SetAlCapNameFnt{\small}
\SetAlCapHSkip{0pt}
\newcommand{\obj}{\mathbf{x}}   

\newcommand{\change}[1]
{#1}
\newcommand{\meas}{\mathbf{b}}
\newcommand{\A}{\mathbf{A}}
\newcommand{\eras}{\mathbf{E}}
\newcommand{\shuttind}{s}
\newcommand{\shuttfunc}{S}
\newcommand{\noise}{\boldsymbol{\eta}}
\newcommand{\perm}{\mathbf{P}}

\newcommand{\object}{\mathbf{x}}
\newcommand{\shutter}{\mathbf{S}}
\newcommand{\vecradflux}{\mathbf{r}}
\newcommand{\sensorradianceflux}{\tilde{r}}
\newcolumntype{M}[1]{>{\centering\arraybackslash}m{#1}}
\usepackage{makecell}

\usepackage{array}
\newcolumntype{L}{>{\centering\arraybackslash}m{2cm}}

\begin{document}
\makeatletter
\let\@authorsaddresses\@empty
\makeatother
\title{Single-shot HDR using conventional image sensor shutter functions and optical randomization}
\thanks{This work has been accepted to \textit{ACM Transactions on Graphics (TOG)} and will be presented at SIGGRAPH 2026. 
This is the author's version of the work. The final version will be published by ACM and available via the ACM Digital Library.}
\author{Xiang Dai}
\affiliation{%
  \institution{UC San Diego}
  \country{United States}}
\email{xidai@ucsd.edu}

\author{Kyrollos Yanny}
\affiliation{%
  \institution{UC Berkeley}
  \country{United States}
}
\email{kyrollosyanny@gmail.com}

\author{Kristina Monakhova}
\affiliation{%
 \institution{Cornell University}
 \country{United States}}

\author{Nicholas Antipa}
\affiliation{%
  \institution{UC San Diego}
  \country{United States}}


\begin{abstract}
High-dynamic-range (HDR) imaging is an essential technique for overcoming the dynamic range limits of image sensors.
The classic method relies on multiple exposures, which slows capture time, resulting in motion artifacts when imaging dynamic scenes. 
Single-shot HDR imaging alleviates this issue by encoding HDR data into a single exposure, then computationally recovering it. 
Many established methods use strong image priors to recover improperly exposed image detail. 
These approaches struggle with extended highlight regions. 
We utilize the \textit{global reset release} (GRR) shutter mode of an off-the-shelf sensor. 
GRR shutter mode applies a longer exposure time to rows closer to the bottom of the sensor. 
We use optics that relay a randomly permuted (shuffled) image onto the sensor, effectively creating spatially randomized exposures across the scene. 
The exposure diversity allows us to recover HDR data by solving an optimization problem with a simple total variation image prior. 
In simulation, we demonstrate that our method outperforms other single-shot methods when many sensor pixels are saturated (10$\%$ or more), and is competitive a modest saturation (1$\%$). 
Finally, we demonstrate a physical lab prototype that uses an off-the-shelf random fiber bundle for the optical shuffling. 
The fiber bundle is coupled to a low-cost commercial sensor operating in GRR shutter mode. 
Our prototype achieves a dynamic range of up to 73dB using an 8-bit sensor with 48dB dynamic range. 
\end{abstract}

%
%


%
%


\begin{teaserfigure}\includegraphics[width=1\linewidth]{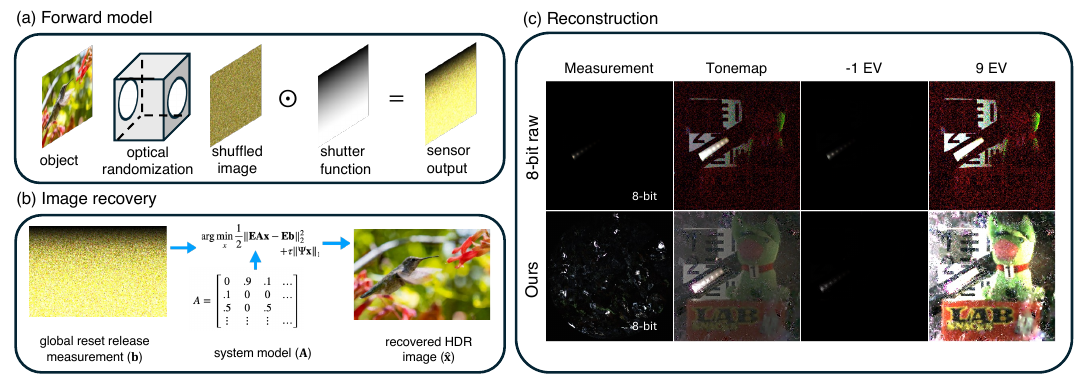}
  \caption{System overview. (a) The scene is optically randomized then recorded using a global reset release (GRR) sensor, which encodes spatially varying exposure via a row-wise integration gradient. (b) We recover an HDR image by solving a simple inverse problem: we ignore saturated or underexposed pixels in the data fidelity loss term--using erasure matrix $\mathbf{E}$--and estimate them using a total variation prior. (c) Our reconstruction from a single 8-bit measurement captures detail across deep shadows and extreme highlights. (Lab Snacks\textsuperscript{\textregistered} is a registered trademark of Thorlabs, Inc.)}
  \label{fig:teaser}
\end{teaserfigure}

\maketitle

\section{Introduction}
Digital image sensors operate with finite dynamic range due to practical limitations such as noise and analog-to-digital converter saturation.  
This limits the image luminance range that can be measured, resulting in loss of detail in dark shadows or bright highlights when both are present in a scene. 
High dynamic range (HDR) imaging is a technique that expands a sensor's detectable luminance range beyond its native capability, improving shadow and highlight performance.
HDR imaging is important in various fields, including automotive~\cite{dufaux2016high}, medical imaging~\cite{seetzen2023high}, and commercial photography~\cite{10.1145/258734.258884}.

A standard method of extending dynamic range is to capture multiple frames with differing exposure times, then combine them in software to create a single HDR image~\cite{10.1145/258734.258884,hasinoff2016burst}.
However, any movement of the sensor or scene while capturing the exposure series leads to artifacts in the HDR image. 
Single-shot HDR methods aim to compute the HDR image from a single sensor frame. 
This improves temporal resolution compared to multi-shot techniques, reducing the impact of motion on the final result.  
However, single-shot HDR is challenging because it often requires solving an ill-posed inverse problem. 

Several single-shot HDR methods have been proposed, which can be categorized into three groups: priors, point spread function (PSF) design, and spatially varying exposure. 
Prior-driven approaches involve capturing a single low dynamic range (LDR) measurement and then using image priors to predict the values of saturated pixels. 
However, this approach cannot accurately reconstruct large, contiguous saturation regions and is prone to hallucinating features that may not be present in the scene~\cite{eilertsen2017hdr, lee2018deep, liu2020single}.
PSF design optimizes the imaging system's PSF to disperse the highlight energy to adjacent pixels or create multiple sub-images with different exposures, forming a better-posed inverse problem. 
These methods also encounter difficulties with large highlight regions, often necessitating strong, learned priors~\cite{metzler2020deep, sun2020learning}. 
Spatially varying exposure methods use optical filters to attenuate light differently at each pixel or manipulate the sensor architecture to create diverse exposure times, then reconstruct the HDR image in a manner similar to demosaicing. 
\change{Approaches that use optical filters to cover the sensor can suffer from crosstalk between adjacent pixels, as discussed in~\cite{schoberl2012high}.}
On the other hand, methods involving custom sensors tend to be complex and costly to implement~\cite{nayar2000high,schoberl2012high,cho2014single,gu2010coded}. 

\begin{figure}
  \includegraphics[width=\linewidth]{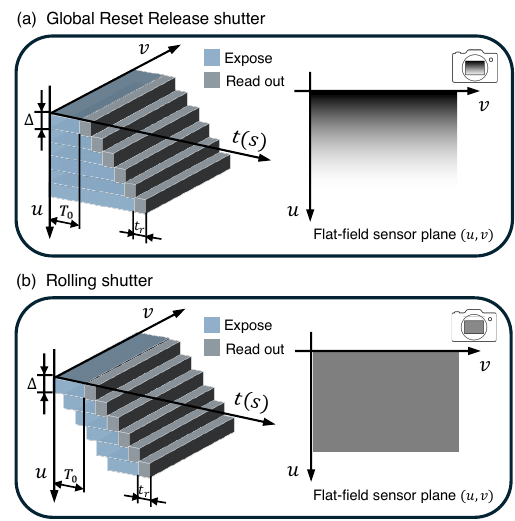}
  \caption{Comparison between GRR and rolling shutter functions. (a) The GRR shutter function starts exposing each row at the same time, so after the initial exposure time ($T_0$), each row ($u$) will have an exposure time that is incremented by the readout time ($t_r$). This results in a row-varying exposure of $T_0 + (u-1)\cdot t_r$ for each column ($v$) within a row. The resulting flat field image (right) has a visible linear gradient. (b) A rolling shutter has the same exposure time $T_0$ and a $t_r$ readout time delay for each row on the sensor. The resulting flat field image (right) is constant.}
  \label{fig: Shutter function comparison}
\end{figure}

Our work utilizes the \textit{global reset release} (GRR) mode of an off-the-shelf rolling shutter sensor, pairing it with optical randomization to generate spatially varying random exposure times across the field-of-view (FoV). We use a simple regularized inverse problem to recover an HDR image from a single measurement (see Fig.~\ref{fig:teaser}).
GRR shutter mode (Fig.~\ref{fig: Shutter function comparison}a) uses the same sequential read architecture as the rolling shutter (Fig.~\ref{fig: Shutter function comparison}b), but each row starts exposing simultaneously.
As a result, each row's exposure time is slightly greater than the row above, resulting in an exposure time gradient across the sensor. 
This sensor mode is intended to be used with a flash triggered during the first $T_0$ seconds when all rows are receptive to light to approximate global shutter. 
Without the flash, the differing exposure times result in a strong exposure gradient (Fig.~\ref{fig: Shutter function comparison}a) in the final image, underexposing the top rows and overexposing those at the bottom. 
To avoid this exposure gradient, we spatially randomize the image before it is measured with a GRR sensor. By mapping each scene point to a pseudorandom pixel, it experiences a pseudorandom exposure time, extending the measurement's dynamic range. 
This randomness means that large highlight or shadow regions contain a mix of well- and poorly-exposed samples. 
We recover the HDR image using standard inverse methods, with only a lightweight total variation prior employed for estimating poorly exposed pixels from their properly exposed neighbors. 
Using this system, we can recover an HDR image with a high saturation rate from a single LDR image.

We validate our method in simulation, comparing it with alternative optical designs, sensor models, and single-shot HDR approaches to show that it is better posed and can successfully recover challenging HDR scenes with dense highlights without fine-tuning the exposure times.
We also demonstrate a prototype using a commercial sensor with the GRR shutter function and a random fiber bundle to verify the concept experimentally. 
Our system can reconstruct an HDR image with a dynamic range of up to 73 dB using an 8-bit (48 dB) input measurement. 

Specifically, we make the following contributions: 
\begin{itemize}
    \item We introduce a novel architecture for single-shot HDR that realizes random exposure times by combining spatially randomizing optics with a GRR sensor.
    \item We compare our method against existing single-shot HDR techniques and demonstrate superior performance across a variety of scenes, including those with extended highlights. Even under high saturation levels (up to 10$\%$), our method consistently outperforms prior approaches across multiple HDR quality metrics.
    \item We demonstrate a prototype with a random fiber bundle and GRR sensor, which achieves dynamic range up to 73 dB from a single 8-bit measurement.
\end{itemize}

\section{Related Work}
\noindent \textbf{Multi-shot HDR imaging.} 
\change{The classic way to recover an HDR image is to capture exposure brackets of LDR images and synthesize them together~\cite{10.1145/258734.258884,mann1994beingundigital,mertens2009exposure,10.1145/1401132.1401170, hasinoff2010noise}. This problem is well studied and has been used in mobile photography by quickly capturing bursts of photos~\cite{hasinoff2016burst,mildenhall2018burst}.}
However, these methods suffer from ghost artifacts due to motion blur in fast-moving scenes. 
\change{Several approaches have been proposed to address motion blur in HDR images, including increasing the ISO setting (i.e., gain) to reduce exposure time~\cite{akyuz2007noise,akyuz2020deep}}, as well as calibrating the camera location and average radiance of the motion-affected region~\cite{gallo2009artifact}. 
More recently, learning-based methods such as the conditional diffusion models~\cite{yan2023toward} and CNNs~\cite{prabhakar2020towards} have been used to address ghosting in HDR images. 
However, these methods are significantly influenced by the quality of the input LDR images. 
Our method encodes different exposure times within a single LDR to reconstruct an HDR image instead of using multiple exposure times, resulting in fewer ghosting artifacts.

\noindent \textbf{PSF engineering.} 
Several single-shot HDR methods implement PSF engineering to spread the energy from bright scene points onto many sensor pixels, thereby providing enough unsaturated measurements to facilitate computational recovery. 
Different optical element designs such as a star filter~\cite{5995335}, multifunctional metasuraces~\cite{brookshire2024metahdr}, and microlens arrays~\cite{cha2023microlens} can modulate the PSF of the optical system to a pattern which will help recover the highlight information in the LDR image and extend the dynamic range. 
Recently, end-to-end methods have been proposed to optimize the PSF of diffractive optical elements in tandem with the parameters of an HDR recovery algorithm~\cite{metzler2020deep, sun2020learning}. 
However, these methods are often optimized for sparse highlights and struggle with large bright patches in the scene.
Our method has random exposure times across the scene, which ensures \textit{some} valid pixels are present extended highlight patches, making the inverse problem of HDR recovery better posed. 
Therefore, we find that our method better handles scenes with dense highlights as compared to PSF engineering approaches.

\noindent \textbf{Spatially varying exposure.} Spatially varying exposure methods use either optical elements or sensor architecture to introduce differing exposure values across the scene. This exposure variation allows computational recovery of an HDR image from a single LDR measurement.
One common approach is to introduce optical filters. 
A conventional camera with a color filter has been shown to extend the dynamic range by utilizing the attenuation of the spectral transmittance of the filters ~\cite{hirakawa2011single}; the reported dynamic range extension was modest.
Another method is to add a neutral density filter array onto the sensor, providing pixel-wise exposure variation ~\cite{aguerrebere2014single, nayar2000high, xu2021deep,schoberl2012high}. 
The classic method creates a superpixel covered by a 2$\times$2 neutral density array~\cite{nayar2000high}, which only has four different exposures and a fixed pattern, and reduces spatial resolution. 

Introducing more exposure times in a single frame on the sensor is an alternate way to have spatially varying exposures.
Dual exposure pixels use a pixel architecture that implements two pre-determined exposure times at each pixel~\cite{an2017single, aguerrebere2014single, gu2010coded}.
To achieve a more comprehensive pixel-wise exposure pattern, several learning-based methods optimize the time-varying exposure design~\cite{9887786, neuralsensors2020, kim2023joint}. 
Also, some specially designed sensors, such as MantissaCam~\cite{9887659}, can extend the sensor dynamic range by using a special analog-to-digital converter (ADC) which wraps values above the ADC limit rather than clipping. 
Event sensors with learning-based methods can reconstruct high-speed HDR video. ~\cite{rebecq2019high,zou2021learning}
Sensor engineering adds significant flexibility to the design; however, customized sensors are expensive and difficult to manufacture at scale. 
Our method achieves spatially varying exposure using off-the-shelf optics and sensors.

\noindent \textbf{Hallucinating HDR via a single LDR image.} A significant amount of work has been done to recover HDR content from an LDR image using deep learning methods ~\cite{eilertsen2017hdr, santos2020single, prabhakar2020towards, wu2022litmnet, li2019hdrnet,an2017single,niu2021hdr}. 
While these methods produce a plausible reconstruction, the complete reconstruction may include inaccuracies when the saturation region is large due to reliance on strong image priors to inpaint these regions. 
Our work uses a more informative LDR measurement approach, thereby drastically reducing the strength of the required image priors used in HDR recovery.

\section{Methods} \label{sec:methods}
This section reviews the system model and illustrates that spatial randomization with the GRR shutter function can provide pseudorandom exposure times in a single frame and formulate a better-posed inpainting problem. 
Given these random exposure measurements, we treat the saturated pixels as erasures and use an inverse problem to inpaint the missing areas, including saturated and underexposed pixels.

\subsection{System Modeling}

The system model includes optical and sensor components.
We start with the sensor model, which takes an input $\sensorradianceflux[u,v]$, which is proportional to radiant flux in J/s, at pixel $(u,v)$ on the sensor plane.
Assuming $\sensorradianceflux$ does not depend on time, the energy deposited at each pixel, $E[u,v]$, is 
\begin{equation}
    \begin{split}\label{eq:sensorradianceflux}
    E[u,v] = \Delta t &\sum_t \shuttind[t|u,v]\sensorradianceflux[u,v]\\
    &=\sensorradianceflux[u,v] \shuttfunc[u,v],
    \end{split}
\end{equation}

\noindent where $\shuttind[t|u,v]$ is a discrete indicator that is 1 for time points when a pixel is receiving photons and 0 otherwise, and $\Delta t$ is the time between each time point in $\shuttind$. 
The total exposure time at each pixel is $\shuttfunc[u,v] = \Delta t \sum_t \shuttind[t|u,v]$. 
Note that we have assumed that the scene and optical system are not time-varying, causing the energy to be approximated as a point-wise multiplication.
The sensor is $i$ pixels wide and $j$ pixels tall, so that $u= \{1,2,\cdots,i\}$ and $v=\{1,2,\cdots,j\}$.  
Equation~\eqref{eq:sensorradianceflux} can be further expressed as matrix multiplication 
\begin{equation}\label{eq:matrixradiantflux}
    \mathbf{E}=\shutter \vecradflux{,}
\end{equation}
 where $\shutter = \mbox{diag}(\mbox{vec}(S[u,v]_{1\leq u \leq i,1\leq v \leq j}))$ is a diagonal matrix and $\vecradflux=\mbox{vec} (\sensorradianceflux[u,v]_{1\leq u \leq i,1\leq v \leq j})$ is a vector.  

The resulting energy at each pixel $(u,v)$ is quantized by the analog-to-digital converter (ADC).
We model the camera response as linear within the ADC range and clipped outside, using the function \( Q(\cdot) = \max(0, \min(2^B - 1, \cdot)) \), where \( B \) denotes the bit depth of the sensor.
Within the ADC range, the difference between the actual quantized signal and the linear approximation is assumed to be an additive noise term $\noise$, including sensor read noise and quantization error.  
Our final sensor model is given by
\begin{equation}\label{eq:sensor_model}
    \meas = Q\{\shutter \vecradflux + \noise\}{.}
\end{equation}

\noindent Variable $\vecradflux$ represents the radiant flux arriving at the sensor inside the camera body. As a result, it is a linear function of the scene radiant flux, $\object$. We denote the linear mapping from the world to the sensor through the optical system as matrix $\perm$, leading to the final forward model expression
\begin{equation}
    \meas = Q\{\shutter \perm \object + \noise\}{.}
\end{equation}

\noindent Next, we model $\shutter$ for the GRR shutter function and argue that when $\perm$ is a random permutation matrix, an HDR scene $\object$ can be estimated from LDR measurement, $\meas$. 

\subsection{GRR shutter function}\label{sec:shutterfunction}
\begin{figure}
    \centering
    \includegraphics{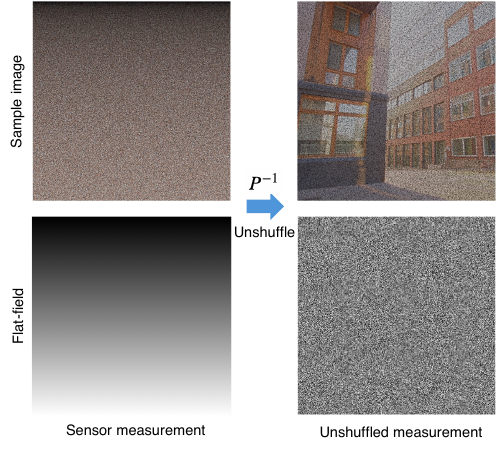}
    \caption{Examples of sensor measurements from our system (left) are randomly permuted and have a linear exposure gradient. Measurements can be unshuffled by the matrix $P^{-1}$ to visualize the scene contents and the pixel-wise random exposure (right). }
    \label{fig:pixel shutter for better visualization}
\end{figure}

\change{The GRR shutter function is designed to mimic global shutter imaging with a low-cost sequential rolling shutter read architecture. Here, we instead utilize it to introduce exposure diversity.}
During the recording of a single frame with a sensor in GRR shutter mode, all sensor pixels are initially receptive to photons. 
After $T_0$ seconds, the first row read process begins, and it ceases to record light. 
The read process takes $t_r$ seconds, after which the next row is read out. 
The remaining rows are read in the same manner, sequentially from the top to bottom. 
Therefore, the total exposure time at each pixel is
\begin{equation}
\label{shutter_equation}
    S[u,v] = T_0+t_r\cdot[u-1]{.}
\end{equation}
\noindent \change{Note that both $t_r$ and $T_0$, and therefore the dynamic range, are controllable. The read time, $t_r$, is determined by the sensor clock speed, and the exposure time, $T_0$, is determined by the user.}
The maximum exposure time is $T_0+(i-1)t_r$, and the minimum is $T_0$. A commercial sensor typically has thousands of rows, so the range of exposures in a single frame can span three orders of magnitude. 
When coupled with a conventional lens, this results in a strong exposure gradient across the image. 
For example, Fig. ~\ref{fig:pixel shutter for better visualization} bottom left shows the measurement of a flat-field (pure white) scene. 



However, utilizing the GRR exposure gradient to improve dynamic range when imaging with a lens will not work for general scenes; it is only feasible for scenes with higher energy at the top and lower energy at the bottom. 
For general scenes, GRR often results in large, contiguous regions being either over- or under-exposed.
Fig.~\ref{fig:meas compare} (GRR shutter) shows this effect: the bright sky is well exposed at the top of the image, but the dark building is under-exposed (highlighted in yellow). 
Conversely, at the bottom of the frame the dark details are well-exposed and brighter details are over-exposed (shown in red). 
Recovering the lost regions of such an image via inpainting over large, contiguous areas is challenging and ill-posed.
Despite the wide exposure range present in GRR images, the fact that natural images contain large areas of similar brightness means that recovering missing pixels in GRR images is often a similarly difficult problem to inpainting on global shutter images. 
To make the problem of recovering extended dynamic range better posed for natural images, we propose using randomly permuting optics to effectively distribute the GRR exposure gradient randomly throughout the scenes. 


\subsection{Spatial pseudorandomness in optical system}\label{sec:3.3_fibers_equal_random_exposure}

\noindent To alleviate this dependency on the scene energy distribution while using the GRR shutter function, we design an optical system that maps the scene's radiant flux pseudorandomly onto the sensor.
We start by assuming $\perm$ is a pseudorandom permutation matrix. 
That is, it shuffles (randomly reorders) the scene vector $\object$, mapping a single scene point to a single pseudorandom sensor pixel. 
When the shuffled scene is measured with the GRR sensor exposure process, the result is effectively spatially random exposure times. 
To see why, we de-shuffle the measurement by applying $\perm^{-1}$ to $\meas$: 

\begin{equation}
\begin{split}\label{eq:permcommuteq}
\perm^{-1}\meas &= \perm^{-1}Q\{\shutter \perm \object\}\\
                &= Q\{\perm^{T}\shutter \perm \object\}{.}
\end{split}
\end{equation}

\begin{figure*}
  \includegraphics[width=\linewidth]{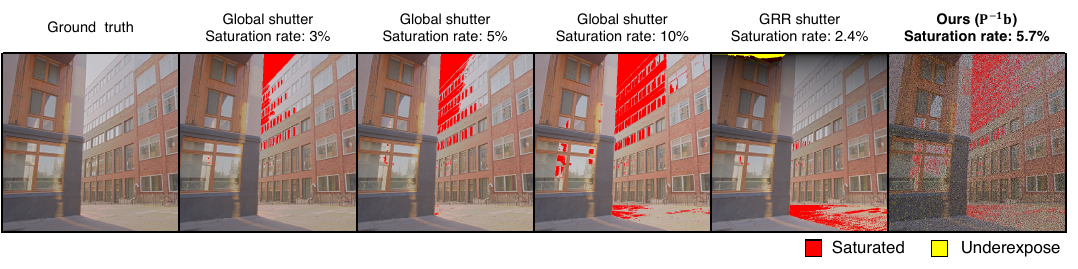}
  \caption{\change{Simulation measurements comparison after apply $\perm^{-1}$ to measurement $\meas$.} We compare the tonemapped HDR image generated by the multi-shot method to measurements using different forward models, including the global shutter with different saturation settings, GRR shutter function, and our method. The exposure range of the GRR setting is designed to cover the exposure times of the input LDR image list. The saturated and underexposed clipped pixels are labeled in red and yellow, respectively and the saturation rates are labeled. Our measurement has more erasure than the GRR shutter function only, but the erasures are better distributed.}
  \label{fig:meas compare}
\end{figure*}

\noindent Because $\perm$ represents a unitary bijective transform (i.e., each row contains exactly one nonzero) and $Q\left\{\cdot \right\}$ is a pointwise nonlinear function, we can exchange the order of $\perm^{-1}$ and $Q\{\cdot\}$.  
The transpose of the permutation matrix maps the shuffled points back to their original location in the scene.
The new measurement $\hat{\meas} = \perm^{-1}\meas$ can be expressed as 
\begin{equation}
    \hat{\meas} = Q\{\boldsymbol{\Lambda}\object\},
\end{equation}
where the $\boldsymbol{\Lambda} = \perm^T\shutter\perm$ is a diagonal matrix with randomly reordered version of $\mbox{vec}(\shutter)$ on the diagonal.
We include a detailed proof in the appendix~\ref{section:appendixmath}. 
The result is that this system effectively applies a \change{spatially} random exposure to the scene prior to quantization,
\change{
as illustrated in Fig.~\ref{fig:pixel shutter for better visualization}.
This observation is central to our design motivation, as spatially varying exposure is known to extend dynamic range ~\cite{nayar2000high}. 

Note that our system's spatially varying exposure values can be controlled by the sensor line time $t_r$ and exposure time $T_0$, whereas focal plane ND filter arrays cannot
}.
Compared to the GRR shutter applied without shuffling (e.g., with a lens), our method has a higher saturation rate since the random exposure time expands the overall dynamic range of a scene. This effect is shown in Fig.~\ref{fig:meas compare}. 
However, the clusters of saturated pixels are smaller and more evenly distributed because many valid pixels have been inserted. This allows recovery of clipped pixels using relatively lightweight image priors such as total variation (TV). 

\subsection{HDR recovery}
HDR recovery from a single LDR measurement is an ill-posed inverse problem because information will always be lost at saturated pixels.
To formulate our HDR recovery problem, we treat saturated pixels as erasures in the measurement. 
We model this using a measurement-dependent diagonal matrix $\mathbf{E}=\mbox{diag}(\epsilon_k)$ where $\epsilon_k$ is the indicator

\begin{equation*}    
    \epsilon_k=    
    \begin{cases}
        1 & \meas_k < 2^B - 1 \\
        0 & \meas_k = 2^B-1.
    \end{cases}
\end{equation*}

\noindent We recover the HDR image by solving a simple inverse problem
\begin{equation} \label{eq:reconstruction}
    \hat \object=\arg \min_\obj \frac{1}{2}\|\eras \A \obj-\eras \meas\|^2+\tau \|\Psi\obj\|_{\mbox{TV}},
\end{equation}
where $\mathbf{A} = \mathbf{SP}$ is the system matrix including the shutter function matrix and optics, $\tau$ is a nonnegative tuning parameter. 
Here, $\Psi$ maps RGB image $\obj$ to YCbCr color space and $\|\Psi \obj \|_{\mbox{TV}}$ is the 2D anisotropic total variation semi-norm of each channel ~\cite{rudin1992nonlinear}. 
We perform reconstruction using FISTA ~\cite{beck2009fast} with the parallel proximal approximation to TV denoising~\cite{kamilov2016parallel}. 
Note that by multiplying both $\mathbf{A}\obj$ and $\meas$ by $\eras$, the data fidelity loss, and hence the gradient, for saturated pixels is zero. 
During reconstruction, this means that only the prior term $\| \Psi \obj\|_{\mbox{TV}}$ is used to effectively inpaint the lost pixels based on the values of neighboring valid pixels. 

We observe that the unique combination of the GRR shutter function and randomizing optics leads to a better posed problem than directly inpainting the raw camera measurements 
, reducing reliance on strong priors.
We demonstrate the property by an ablation test using different combinations of optical system and sensor shutter functions in the following section.

\section{Simulation Results}
In this section, we perform an ablation study using different combinations of sensor shutter functions and optical systems. We find that the \textit{combination} of randomization and GRR shutter significantly outperforms the use of one or the other alone. 
Then, we compare our method to a selection of published state-of-the-art single-shot methods, including Deep Optics HDR~\cite{metzler2020deep}, HDR-CNN~\cite{eilertsen2017hdr}, and spatially varying exposure~\cite{nayar2000high}. We find that our method handles a wider range of scenes and saturation rates without adapting exposure values per-scene. 
However, most published methods work better in their specific design conditions than our method. 

\subsection{Metrics and dataset preparation}\label{sec:metrics and dataset preparation}
\change{We preprocess ground truth and reconstructed HDR images using the PU21 transformation\footnote{This step improves agreement between pixel-wise loss functions and human perception for HDR images}~\cite{azimi2021pu21} before evaluating HDR performance using PSNR, $\mbox{PSNR}_\gamma$, SSIM, MSSSIM, VSI, and FSIM.
Unless otherwise specified, all input images are normalized by their maximum values to make sure the range is between 0 and 1 before evaluation, reflecting our emphasis on accurate highlight reconstruction, which is critical for HDR quality.
We use the PU21 default settings: peak luminance 100 cd/m\textsuperscript{2}, contrast ratio 1000:1, and an ambient light level of 10 cd/m\textsuperscript{2}.}

\change{We also include the perceptual visual metric HDR-VDP-3 (Q-score and $Q_{\mathrm{JOD}}$)~\cite{mantiuk2023hdr} (peak luminance 100 cd/m\textsuperscript{2}). 
The Q-score is a comparative perceptual quality metric ranging from 0 to 10, where higher values indicate better perceived quality.
$Q_{\mathrm{JOD}}$ is a variant of the Q-score that quantifies perceptual differences on a Just-Objectionable-Differences (JOD) scale.
Specifically, a decrease of 1 point corresponds to approximately 75\% of the general population perceiving a loss in quality.
In our experiments, HDR-VDP-3 is used in side-by-side mode with a visual acuity setting of 30 pixels per degree to match our experimental conditions.}
We use all 180 images from the SI-HDR dataset~\cite{hanji_mantiuk_eilertsen_hajisharif_unger_2022} as our test scenes. This dataset covers a diverse set of environments, including nighttime and daytime cityscapes, forests, indoor scenes, and more.
Each scene includes five raw images captured using a Canon EOS 5D Mark III, with exposure times logarithmically sampled from 1/8000s to 1/2s.
\change{To generate the ground truth HDR images, we first read the demosaiced images from the .CR files, then crop each to a 3840×3840 region. The cropped images are downsampled to 512×512 using box filtering to reduce aliasing. Finally, we apply MATLAB’s built-in multi-shot HDR generation function~\cite{10.1145/258734.258884,reinhard2020high} to merge the exposures.}

\renewcommand{\arraystretch}{1.5}
\newcolumntype{C}[1]{>{\centering\arraybackslash}m{#1}}

 \begin{figure}
  \includegraphics[width=\linewidth]{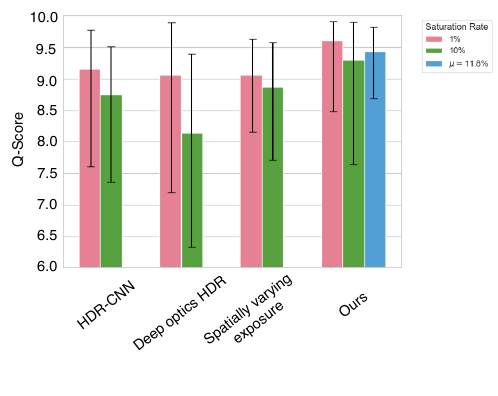}
  \caption{\change{Baseline test comparison Q-score.} Bar plot of Q-score reconstruction using HDR-CNN, Deep Optics HDR, spatially varying exposure, and our method under different saturation settings. The error bar indicates the value from the 5th percentile to the 95th percentile. Our method is additionally tested with fixed gain and shutter function, resulting in an average saturation rate $\mu=11.8\%$. Our method performs well for various scenes in the dataset with either low or high saturation rates.}
  \label{fig:baselinetestplot}
\end{figure}
\begin{figure*}
      \includegraphics[width=\linewidth]{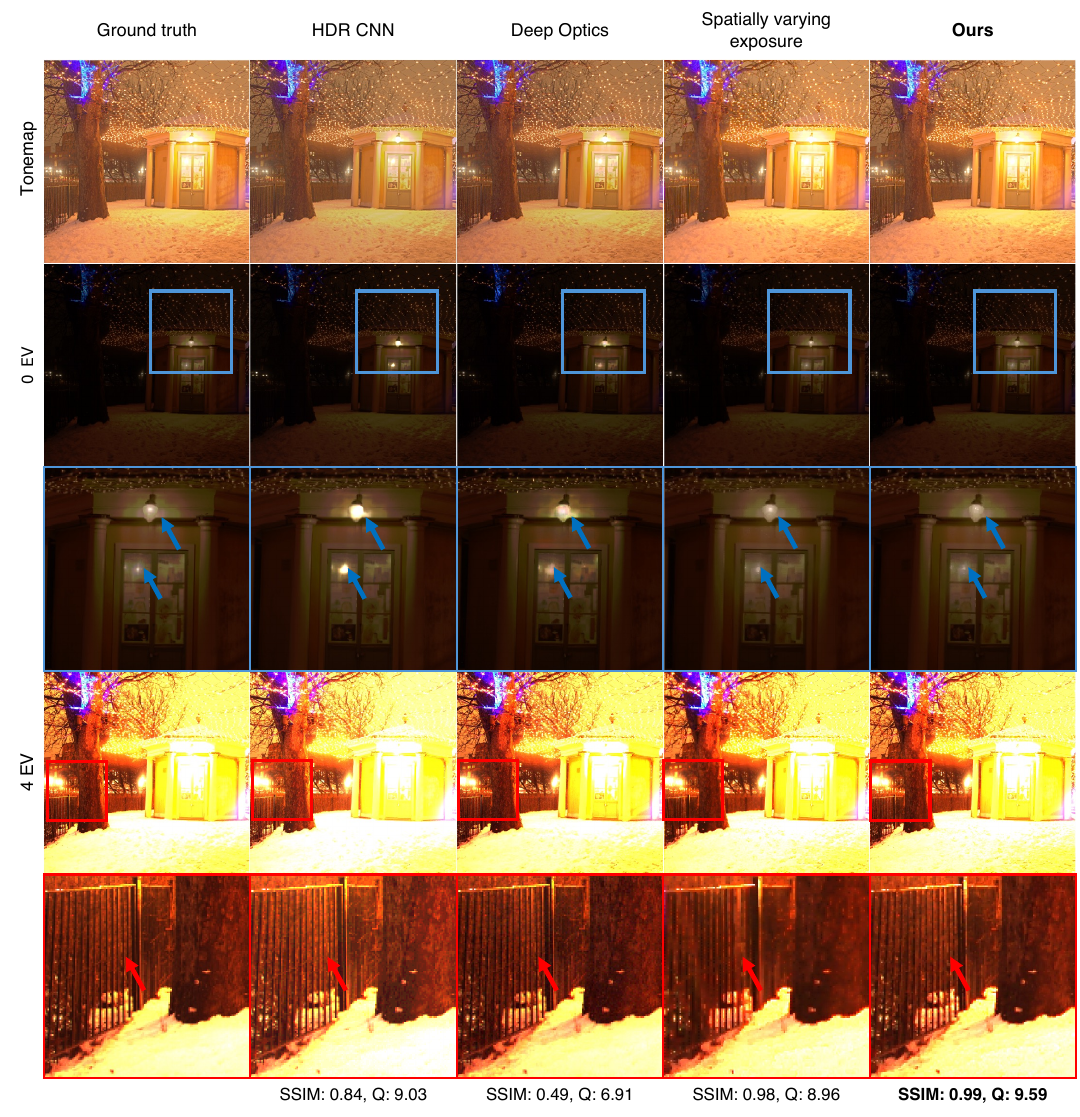}
      \caption{\change{Reconstruction example of $1\%$ saturation rate in the baseline test.} The image includes the tonemap, 0 EV, and 4 EV of the ground truth and the reconstruction of HDR CNN, Deep Optics HDR, spatially varying exposure, and our method. Zoom-ins are displayed under each image to highlight details in the reconstructions. The blue arrow indicates the artifacts in Deep Optics and HDR-CNN in the bright lamp. The red arrow indicates the artifacts in spatially varying exposure in the fence posts. }
      \label{fig:baseline test}
\end{figure*}

\subsection{Baseline test}\label{sec:baseline_4_2}
This section compares our methods with other state-of-the-art single-shot HDR methods, including HDR-CNN~\cite{eilertsen2017hdr}, neutral density (ND) filter array~\cite{nayar2000high}, and Deep Optics HDR~\cite{metzler2020deep} in simulation using the same data and preparation method as in the ablation test.

We use the pre-trained network parameters for HDR-CNN.
For Deep Optics HDR, we simulate the LDR image using the published PSF, then reconstruct the HDR image using the pre-trained network weights. 
The ND filter array comprises tiling a 2x2  filter pattern with transmittances 1/128 (OD 2.1), 1/32 (OD 1.5), 1/8 (OD 0.9), and 1/2 (OD 0.3), clockwise from the upper left.
Instead of using bicubic interpolation to estimate saturated pixels, as was done in the original publication~\cite{nayar2000high}, we apply our erasure-based TV-constrained inverse solver. 
We test each method at both 1\% and 10\% saturation. 
For our method, this requires adjusting the shutter function on a per-scene basis. 
Because we do not do this in practice, we also include an analysis of our system with a static shutter function that covers the exposure range of the ground truth exposure bracket set.
This results in scene-dependent saturation rates with a mean of 11.8\%. 
The results of the baseline tests are shown in Fig.~\ref{fig:baselinetestplot}, and the full metrics analysis table is in appendix Table.~\ref{tab:baseline test} in which all the data is normalized by its maximum values. 

Our method performs better in mean value and standard deviation of all the metrics than the 2$\times$2 ND filter array with $1\%$ and $10\%$ saturation rates.  
\change{Therefore, some diversity in exposure with random distribution improves the inpainting problem.} 
\change{HDR-CNN and Deep Optics HDR both exhibit high variance when the data are normalized by their maximum values.
This is primarily because these methods hallucinate information in saturated regions, leading to inaccurate highlight reconstruction, and therefore incorrect dynamic range estimation. 
This contributes to higher variance across metrics~\cite{eilertsen2017hdr}.
In contrast, spatially varying exposure and our method achieve higher data fidelity by capturing multiple exposures within a single shot. 
This enables more accurate highlight recovery and results in lower variance in perceptual metrics such as the Q-score.
} 
We find that both learning-based methods work best for samples with lower saturation rates and without large, contiguous saturated image patches. 
The reconstruction quality drops once the saturation rate increases, especially when dense highlights are present. 
Our method is more robust to higher saturation rates and large highlight regions. 
At an average saturation rate of around 11.8$\%$, our method achieves stable reconstruction quality across all metrics, even without per-scene adjustment of the camera gain or shutter function parameters.
A complete baseline test analysis is available in Table.~\ref{tab:baseline test} in the appendix.

\change{An additional analysis is presented in Appendix Table~\ref{tab:baseline test mean norm} and the corresponding bar plot in Fig.~\ref{fig:baseline test norm by mean}, using the same evaluation setting but with both the reconstruction and ground truth images normalized by their mean values.
Under this normalization, the variance of both learning-based methods decreases, as the impact of highlight reconstruction errors is diluted by the larger number of well-exposed pixels. 
Despite this, our method still achieves the highest mean performance across different saturation rates in most of the metrics.
However, since our primary goal is to recover highlight information accurately, we adopt maximum-value normalization as the main preprocessing step throughout our evaluation.}

\noindent In Fig.~\ref{fig:baseline test}, we display an example HDR reconstruction using each baseline method with $1\%$ sensor saturation. 
We observe that highlights estimated by learning-based methods are prone to artifacts (see blue arrows in the 0 EV views outlined in blue).
In contrast, the ND filter array and our method perform well in highlights. 
Overall, our method performs better than the ND filter array in reconstructing shadows. Deep Optics HDR and HDR-CNN perform the best in the shadows. For example, see the details in the fence called out by the red arrow in the 4 EV inset. 
\subsection{Synthetic ablation test}
In this section, we implement different combinations of optics, shutter functions, and image priors. 
Our approach comprises an optical randomizer and GRR shutter function. We compare ours to a lens ($\perm=\mathbf I$) paired with both the GRR shutter and global shutter function ($\shutter = 1$). 
All systems use a sensor with $B=8$ and $\sigma = 0.01$ additive Gaussian noise. 
For each simulated hardware combination, we reconstruct HDR images with and without TV as the prior in the reconstruction.
The \textit{combination} of optical randomization and GRR shutter function performs better than either alone. 

We begin with the global shutter as the sensor and a camera lens as an optical system. 
We adjust the system's exposure time based on each input to achieve a saturation rate of 3$\%$, 5$\%$, and 10$\%$. 
Example raw measurements are shown in Fig.~\ref{fig:meas compare}. This confirms the expected behavior wherein one must choose to preserve highlights at the expense of shadows or vice versa. 
Since the exposure time is constant across the sensor, optical randomization would not change the result, so we do not include it.

Next, we examine the GRR shutter function combined with a camera lens. The exposure time range of $\shutter$ is chosen such that it covers the exposure range of the ground truth exposure bracket set: $\min(S[u])=1/8000s$ and $\max(S[u])=1/2s$.
Measurement examples are shown in Fig.~\ref{fig:meas compare} with saturated pixels highlighted in red and underexposed pixels in yellow.
\change{This test allows the saturation rate of the measurement to float based on the scene. Therefore, we repeat the same experiment but vary the input peak luminance to achieve specific saturation rates of 3$\%$, 5$\%$, and 10$\%$. Both methods are evaluated using TV as the reconstruction prior, with quantitative results shown in Appendix Table \ref{tab:ablation test}.}

\begin{figure*}
    \centering
    \includegraphics{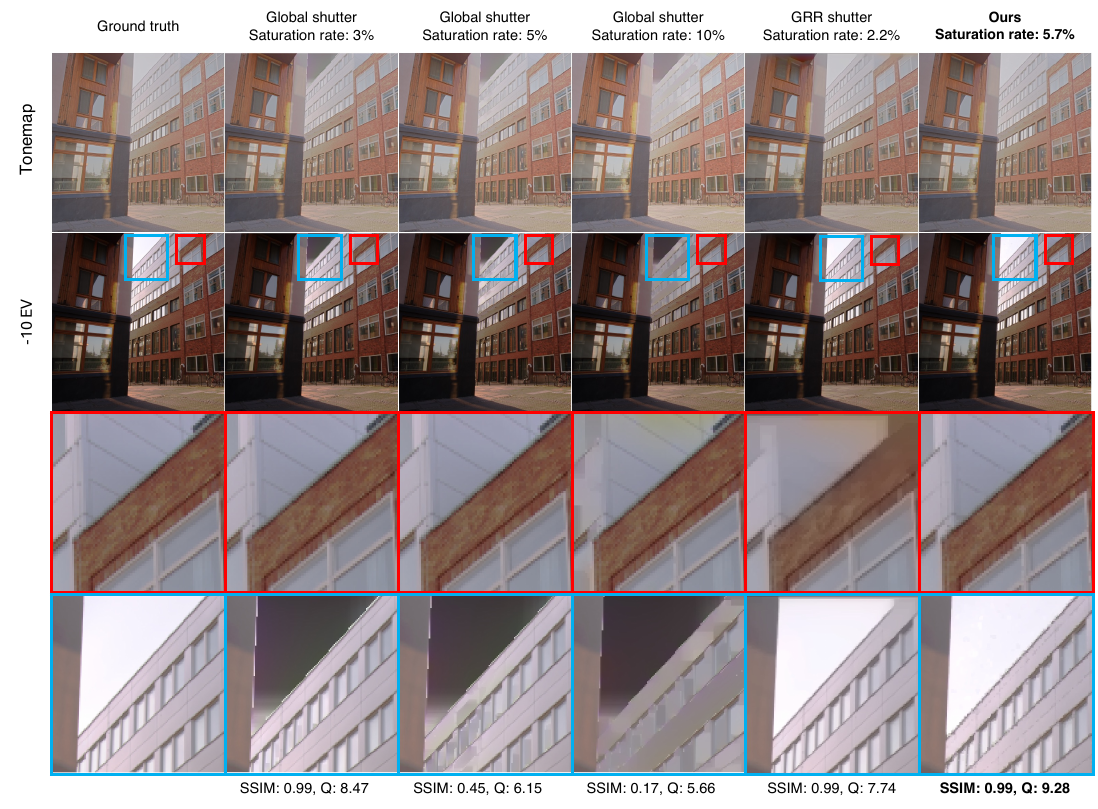}
    \caption{\change{Simulation results.} We compare the ground truth generated by the multi-shot HDR to the reconstruction results using various forward models with a prior. These models include the global shutter function with lens and 3$\%$, 5$\%$, and 10$\%$ saturation rates, the GRR shutter function with lens, and our method. The GRR shutter function range is adjusted to cover the input exposure time range of the input LDR image list. We visualize the reconstruction results using a tone map and -10 EV in the first two rows. Additionally, we zoom in on two regions depicted in red and blue on the -10 EV images to show the detailed reconstruction quality. Our method performs best over different combinations, even with a relatively high saturation rate.}
    \label{fig:sim_compare}
\end{figure*}

Our method shuffles the image and implements the same exposure setting as the GRR shutter function with the camera lens. \change{In Figure \ref{fig:meas compare}, we show $\hat{\meas} = \perm^{-1}\meas$, which clearly demonstrates the randomly distributed exposures and saturated pixels predicted in Sec. \ref{sec:3.3_fibers_equal_random_exposure}. The lens-only systems exhibit large contiguous regions of saturation, which are much harder to inpaint}. The simulated measurements from all hardware combinations are used as input to Eq.~\eqref{eq:reconstruction} with $\tau=0$ (TV prior inactive) and with $\tau >0$ (TV prior active). We score the results using each metric mentioned above and report the mean and standard deviation in Appendix Table \ref{tab:ablation test}. \change{Visual results are shown in in Fig.~\ref{fig:sim_compare}.}
We observe a clear trend that the reconstruction results are better when a prior is included over all the methods.
Higher saturation rates lead to poorer reconstruction results when the global shutter with a camera lens is the system.
With the help of the prior and various exposure times in a single frame, the GRR shutter function with lens has an improved reconstruction result quality.
This is largely driven by scenes in which highlights are concentrated at the top of the scene. 
This results in a high standard deviation in all metrics because many natural scenes do not have a brightness gradient that conveniently aligns with the shutter function.
Our method outperforms the others in mean values and standard deviation, even with the highest average saturation rate. 
Therefore, we conclude that the combination of randomization and GRR shutter is far better than using a lens with the global or GRR shutter.

\section{Hardware demonstration}\label{sec:hardware_demo}
Two essential parts of the design are spatial randomness in the imaging system and the sensor's global reset release shutter function. 
To achieve spatial pseudorandomness, we utilize a random fiber bundle. 
Random fiber bundles comprise many optical fibers that are randomly sorted. 
They are typically used for transporting and homogenizing illumination light. 
We utilize the randomness to provide the image permutation property we need. 
Our fiber bundle contains 846,450 borosilicate multimode fibers with a 25-micron diameter and a numerical aperture of 0.66. 
The scene is imaged onto the input face of the fiber bundle using a Canon 50 mm f/1.4 lens. 
The image projected onto the bundle appears randomized at the output face of the bundle, which is optically relayed using a 4-f system onto a commercial sensor equipped with rolling and GRR shutter modes. The camera is an Allied Vision 1800 U-1240 (Sony IMX 226 chip), with 4024 $\times$ 3036 pixels \change{whose readout time per pixel $t_r$ can be changed by varying the camera clock speed.} A prototype photo is included in Appendix Fig.~\ref{fig:prototype}.

Nonidealities in our hardware system are present, including optical blur and non-one-to-one mapping from the input face to the output of the bundle. 
To capture all of these effects, we experimentally capture the system PSF from each input point across the FoV. 
After calibration, we switch the shutter to GRR mode and capture data from scenes displayed on the calibration OLED TV and from real scenes. To provide a qualitative comparison, we also capture images of the test scenes using exposure bracketing with a Canon 5D Mark II camera. Note that these images are acquired from a different perspective, so the comparison is only qualitative. 




\subsection{Calibration model}
\begin{figure}
  \includegraphics[width=\linewidth]{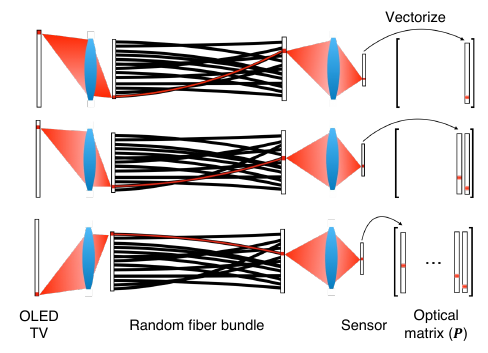}
  \caption{Calibration system overview. Spatial points are displayed and sequentially scanned using an OLED TV as the light source. Then, a camera lens relays the image onto the fiber bundle. At the fiber bundle's right side, another relay image system generates the impulse response onto the sensor. The output image from the sensor is vectorized and stacked horizontally as an optical matrix $\perm$.}
  \label{fig:calibration}
\end{figure}

To calibrate the system’s PSF matrix $\perm$, we displayed a white point source on an OLED TV rather than a structured pattern, as discussed in Appendix Sec.~\ref{sec:HDR ground truth}.
Perhaps unsurprisingly, we found that the fidelity of the PSFs comprising $\perm$ must be quite high, demanding HDR capture of each calibration point. 
Additionally, to prevent $\perm$ from being impractically large, we adopt a sparse matrix representation. 
Here, we describe a calibration approach that achieves both HDR impulse calibration and matrix sparsity \change{as Fig.~\ref{fig:calibration} shows}. 

To find $\perm$, we set the sensor to rolling shutter mode and scan a point source across the TV. 
For exposure time $T$ and a point source at index $k$, the sensor measurement at pixel $(u,v)$ is given by 
\begin{equation}
    \meas[u,v|k,T] = Q\{T\perm[u,v|k]+\noise_{k,T}\}{,}
\end{equation}
where we have absorbed the optical blurring of the main lens and relay system into $\perm$. 
Assuming no saturation, the radiant flux of the impulse response at this pixel location is $\phi_{T} \propto \meas[u,v|k,T]/T$. 
We observe, however, that directly recording each PSF with a single measurement results in significant error due to noise and quantization. 
To reduce these effects, we acquire a set of $m$ measurements, each with a different exposure time, at each calibration location $k$. For each exposure time, we estimate the flux value, storing the series in a vector.

\begin{equation}
    \boldsymbol{\Phi} = [\phi_{T_1}, \phi_{T_2}, \ldots, \phi_{T_m}]^\intercal{.}
\end{equation}

\noindent We also construct an erasure indicator vector $\epsilon$. Element $l$ of $\epsilon$ takes on a value of 1 if pixel $(u,v)$ contains a valid value for exposure time $T_l$ and 0 otherwise. This allows us to estimate the radiant flux at pixel $(u,v)$ given a point source at $k$ as

\begin{equation}
    \hat{\perm}[u,v|k] = \frac{\epsilon^\intercal \boldsymbol{\Phi}}{\mathbf{1}^\intercal \epsilon} {.}
\end{equation}

This calculation is made at all sensor pixels where the following conditions are met. First, the pixel needs to contain at least 3 valid values (i.e., $\mathbf{1}^\intercal \epsilon \geq 3$).   Second, the correlation coefficient between measurements and the exposure times must be greater than 0.95. This step eliminates unresponsive sensor pixels. Lastly, we detect pixels with excessive dark current using dark frames. For any of these invalid pixels, we set their value to zero throughout all of $\perm$ and include their location in $\eras$, effectively discarding them from the inverse problem. The result is a sparse HDR estimate of the $k$-th column of $\perm$. This process is repeated over the FoV until the entire matrix is captured. 

\subsection{Color processing}
To process color images, we subsample the calibration matrix, creating four sub-matrices, one for each subpixel in the sensor's Bayer filter pattern: $\perm_R$, $\perm_{G1}$, $\perm_{G2}$, and $\perm_B$. Similarly, we compute four corresponding shutter matrices by subsampling the shutter function: $\shutter_R$, $\shutter_{G1}$, $\shutter_{G2}$, and $\shutter_B$. To process raw data, we subsample it into four measurements $\meas_R$, $\meas_{G1}$, $\meas_{G2}$, and $\meas_B$, which are vectorized and column stacked. The linear forward model mapping from the RGB scene to the reordered sensor raw is then
\begin{equation*}
    \begin{bmatrix}
        \meas_R\\
        \meas_{G1}\\
        \meas_{G2}\\
        \meas_B
    \end{bmatrix}
    = 
    \begin{bmatrix}
        \shutter_R & 0 & 0 & 0\\
        0 & \shutter_{G1} & 0 & 0\\
        0 & 0 & \shutter_{G2} & 0\\
        0 & 0 & 0& \shutter_B
    \end{bmatrix}
    \begin{bmatrix}
        \perm_R & 0 & 0\\
        0 & \perm_{G1} & 0 \\
        0 & \perm_{G2} & 0\\
        0 & 0 & \perm_{B}
    \end{bmatrix}
    \begin{bmatrix}
        \obj_R\\
        \obj_G\\
        \obj_B
    \end{bmatrix}
    {.}
\end{equation*}

\noindent We then use the inverse problem strategy outlined in Section \ref{sec:methods} to recover the image. We calibrate the system in 8-bit for 400$\times$400 scene points. The resulting calibration matrix is only 940 MB.


\subsection{Quantitative validation}
\begin{figure} 
  \includegraphics[width=\linewidth]{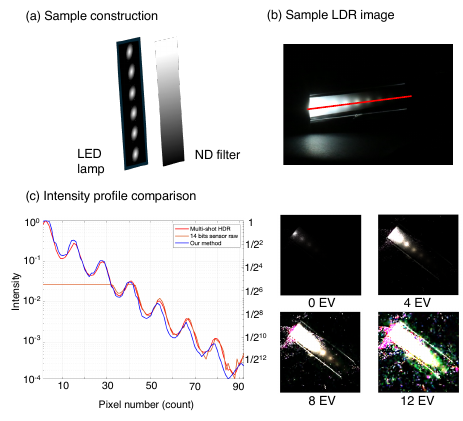}
  \caption{Quantitative validation of the dynamic range extension. (a) Quantitative sample construction. We cover an LED lamp with an ND filter with a 0.4 and 4 OD range. The red line indicates the intensity profile plotted in (c). (b) An LDR image of the sample was taken by a Canon 5D Mark II using a 1/40 shutter function and a 50 mm camera lens. (c) Intensity profile comparison between the multi-shot HDR image and our method. On the left side, the plot of the pixel number against normalized energy on a log scale. On the right, the reconstruction images are displayed using different exposure values. }
  \label{fig:quantitative validation}
\end{figure}
We conduct a quantitative validation to test the effectiveness of our dynamic range extension. 
To make a standard sample, we use a DC-powered LED lamp covered by a continuous neutral density (ND) filter (Thorlabs NDL-25C-4) with an optical density varying linearly between 0.04 to 4.0; this attenuates the LED intensity by between 1 (no ND filter) and 10,000 times. A schematic of our setup is shown in Fig.~\ref{fig:quantitative validation}(a), and a photograph taken with a standard camera is shown in ~\ref{fig:quantitative validation}(b). 
The GRR shutter function we use in our experimental system has a minimum exposure time of $T_0=189 us$.
The readout time is \change{set at} $t_r=51 us$, which, over 3036 rows of pixels, results in a maximum exposure time of $T_0 + 3035t_r = 155$ ms, roughly 820$\times$ higher than $T_0$. 
For brightness analysis, we convert the recovered image from RGB to luminance.
In Fig.~\ref{fig:quantitative validation}(c), the ND filter reconstruction from an 8-bit measurement is displayed with a range of exposure values (EV). 
Note that EV is a $\log$ 2 scale, so an image with $+g$ EV has been multiplied by $2^g$ and clipped before display. 
A standard 8-bit sensor adjusted by +8 EV would effectively be a 1-bit image. 
We observe that our reconstruction adjusted by +8 EV still has detail in the shadows and relatively little noise. 
To compare our method against ground truth, Fig.~\ref{fig:quantitative validation}(c)(left) shows a plot of the reconstructed luminance profile from our method (8-bit measurement) against our ground truth camera (Canon 5D Mark II + 50 mm lens). 
Both 14-bit raw (gold) and multi-shot HDR (red) are plotted. 
Note that the oscillations are caused by the individual LEDs within the lamp. 
The reconstruction's energy attenuation matches the trend of the ND filter specification and the intensity profile from the ground truth multi-shot HDR image.
The 14-bit sensor raw data does not capture the full dynamic range of this scene, and it gets noise when the intensity is low. 
Finally, we reconstruct images from two different LED lamp orientations to eliminate ensure that we did not achieve high dynamic range by aligning the filter gradient with the sensor gradient inadvertently; see Appendix Fig.~\ref{fig:two_orientation_ndfilter}.
From the profile in Fig.\ref{fig:quantitative validation}(c), the maximum and minimum energy attenuation ratio is around 5000 times.
Therefore, the dynamic range of the reconstruction is around 73 dB when using an 8-bit measurement as input, which is higher than the ideal dynamic range of an 8-bit image (48 dB) and higher than the sensor dynamic range in a single frame from its spec sheet (our sensor is specified at 65 dB).
\begin{figure*}
    \centering
    \includegraphics{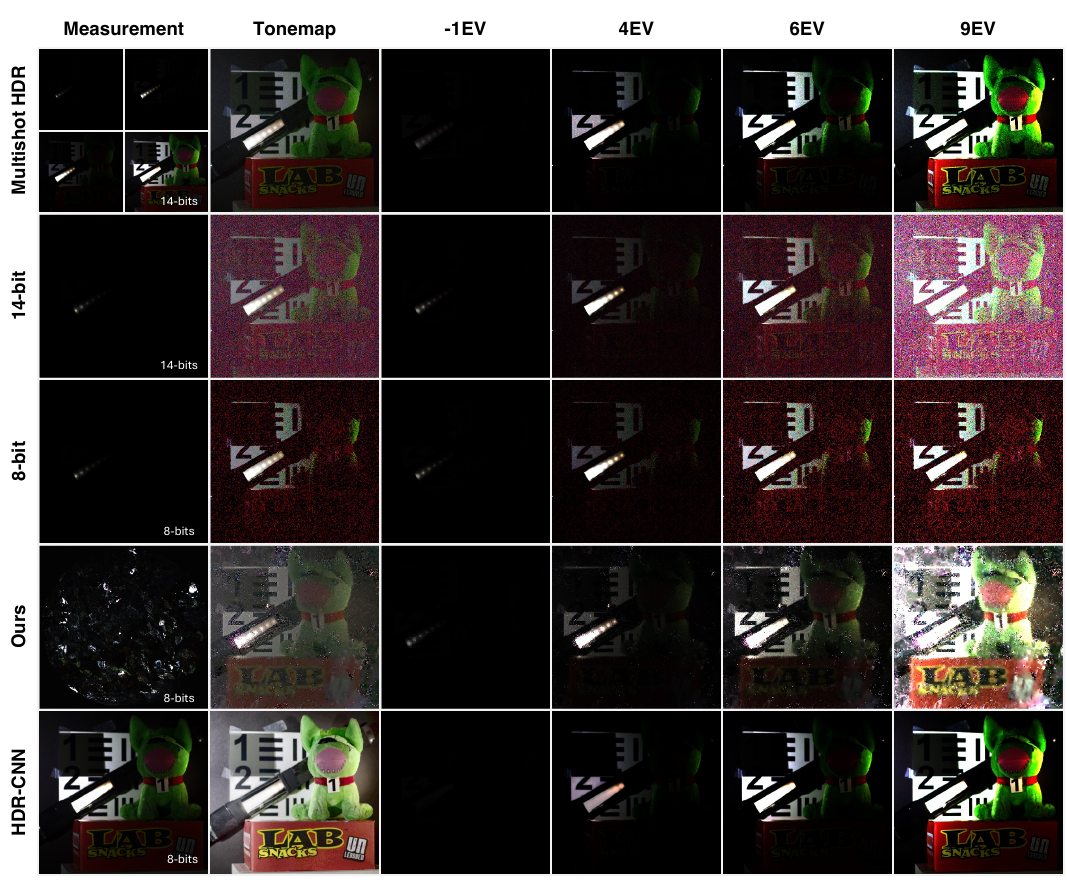}
    \caption{\change{Experimental results.} We qualitatively compare our method against a multishot HDR method, a 14-bit raw image from Canon 5D Mark II, an 8-bit image from a reference camera placed next to our prototype (resulting in slightly different perspectives) and HDR-CNN when input image well exposed the low energy part. The sample contains a continuous neutral density filter (0-2 OD), and we display the methods under different exposure values (EV). Our method can reconstruct information whose dynamic range is close to 14-bit sensor raw data from the commercial camera using an 8-bit LDR input. The sample Lab Snacks\textsuperscript{\textregistered} box used in the experiment is a registered trademark of Thorlabs, Inc.}
    \label{fig: qualitative results of ndfilter with dog}
\end{figure*}
\subsection{Qualitative results}
To qualitatively assess our approach, we image an HDR lab scene using a variety of surface textures and brightness. We make a qualitative comparison to images captured using the ground truth camera. 
Note that the ground truth is taken from a different perspective than our image, so comparisons cannot be made pixel-for-pixel. 
Within the scene, we use a variety of colors and textures, including a toy dog and a USAF 1951 test target. 
To introduce a gradation of highlights, we include the same LED lamp as the previous section, this time attenuated with a linear ND filter with a range of 0 to 2 OD. 
We also mask a square region of the USAF target with a film ND filter with OD 0.6. 
The GRR shutter function has an initial exposure time $T_0=189 us$ and a readout time \change{sets at} $t_r=188 us$ over 3036 rows. 
We display the reconstruction images Fig.~\ref{fig: qualitative results of ndfilter with dog}, with multi-shot HDR reconstruction, 8-bit sensor raw, and 14-bit sensor raw data for visual comparison.
Because of the limited dynamic range, the 8-bit measurement lacks shadow detail entirely when the highlights are not over-exposed, as shown in the +9 EV adjusted image.  
The 14-bit sensor raw data can record all the information, but the shadows contain significant noise. 
Our method successfully recovers the highlight and shadow details with results that are comparable to those of the 14-bit 5D Mark II, albeit with noticeable qualitative differences between the two. 
\change{Note that we also input an 8-bit image where the low-intensity regions are well-exposed into HDR-CNN for comparison. As shown in the bottom row of Fig.~\ref{fig: qualitative results of ndfilter with dog}, HDR-CNN struggles with this scene and fails to recover the highlight details effectively.}
We believe the noise and nonuniformity in our reconstructed image are due to a model mismatch in our system calibration. 
Sources of mismatch could include stray light, light leakage between fibers in the fiber bundle, broken (dark) fibers, and noise in the HDR calibration values. 

Note that the fibers within the bundle have varying transmittance values. 
This leads to a natural question: are we observing dynamic range extension due to the bundle effectively acting as a spatially varying ND filter array? 
To rule this out, we use simulation to compare our method (experimental fiber matrix combined with 8-bit GRR sensor) to a hypothetical system comprising an ND filter array with per-pixel transmittance equal to the row sum of $\perm$ and coupled with an 8-bit global shutter sensor. We find that the fiber-to-fiber transmittance variation is not enough to explain our results, failing completely for the test scene shown in Appendix Fig.~\ref{fig:fiberproperty test} (bottom row).

\section{Discussion}
In summary, our method can extend the dynamic range of an 8-bit image from its native 48 dB to 73 dB, which is close to a 14-bit extension.
Additionally, our method does not require precise control of the saturation rate and can handle various scenes without requiring adjustments to the sensor or the mask in front of it.
In particular, our design introduces a wide range of exposure values into a single shot by combining spatial randomness and the GRR shutter function, making the problem better posed for reconstruction, even when using a weak prior such as total variation.
This section will discuss dynamic range analysis, motion blur, depth sensitivity, limitations, and future work. 
\subsection{Dynamic range analysis}\label{sec:dynamic_range_analysis}
\noindent 
In prior work based on ND filter arrays ~\cite{nayar2000high}, the dynamic range of the system is given by
\begin{equation}
\label{eq:dynamic range equation}
    DR=20\log_{10}\frac{I_{\mathrm{max}}}{I_{\mathrm{min}}}\frac{e_{\mathrm{max}}}{e_{\mathrm{min}}},
\end{equation}
the $I_{\mathrm{max}}$ and $I_{\mathrm{min}}$ are the sensor's maximum and minimum gray level, respectively. Variables $e_{\mathrm{max}}$ and $e_{\mathrm{min}}$ are the maximum and minimum transmittance, respectively, of the filter array.

\change{The dynamic range of our system can be expressed by combining the shutter function Eq.~\eqref{shutter_equation} with Eq.~\eqref{eq:dynamic range equation}:}
\change{
\begin{equation}
\label{eq:dynamic range ours}
DR = 20\log_{10}\frac{I_{\mathrm{max}}}{I_{\mathrm{min}}} + 20\log_{10}\left(1 + \frac{u_{\mathrm{max}} - u_{\mathrm{min}}}{\frac{T_0}{t_r} + u_{\mathrm{min}-1}}\right).
\end{equation}
}
\noindent\change{This expression shows that the dynamic range of our method is governed by the parameters $T_0$ and $t_r$.
Appendix Fig.~\ref{fig:HDR_t0_tr} shows two simulation examples with different dynamic ranges, tested by varying the ratio between $T_0$ and $t_r$.
These examples demonstrate that different scenes may require different exposure settings to achieve optimal performance.
The parameter $T_0$ represents the baseline exposure time and can be directly controlled via the camera’s exposure setting.
The parameter $t_r$ represents the readout time per row in the GRR shutter and can be modified by adjusting the sensor clock speed or equivalent settings.
In our system, $t_r$ is effectively determined by the throughput limit and can also be indirectly influenced by changing the frame rate when the exposure time is fixed.}

\begin{figure}
  \includegraphics[width=\linewidth]{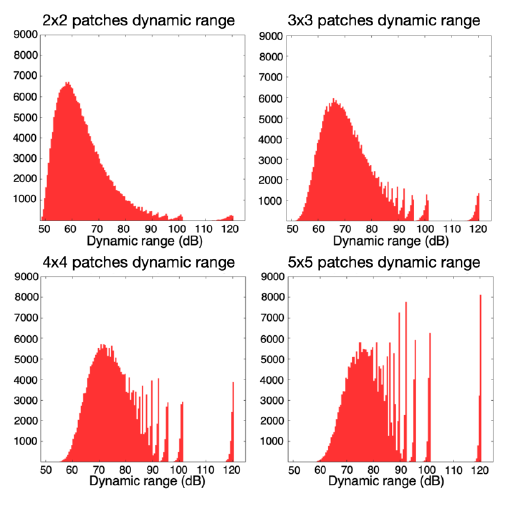}
  \caption{\change{Dynamic range analysis of our system using different patch sizes.} Four histograms of the dynamic range over 2, 3, 4, 5 patches over an image size of 512 by 512 are displayed. The larger patch size is considered the dynamic range over the image distributed at higher values.}
  \label{fig:dynamic range analysis}
\end{figure}

\change{In contrast to methods such as grid-based ND filters, where dynamic range analysis can be conducted using a global equation due to the presence of an orderly superpixel structure, our approach requires a patch-wise analysis.
This is because applying a global maximum-to-minimum exposure ratio across the entire sensor can be misleading in our case:
It is significantly harder to reconstruct a saturated pixel using information from a distant pixel, even if that pixel has a usable exposure.
When exposures are spatially close, a simple regularizer such as total variation (TV) is often sufficient for reconstruction. However, as spatial distance increases, the effectiveness of such priors diminishes.
Instead, we estimate the dynamic range locally across the image using patches of varying sizes, from 2×2 to 5×5 pixels.
For each patch in object space, we compute the dynamic range using Equation~\eqref{eq:dynamic range ours}, based on the exposure values within that patch.
For each patch size, we slide overlapping patches across the entire 512×512 field of view and compute the local dynamic range for each one.
This produces a distribution of dynamic range values across the image, which we summarize using a histogram to visualize how frequently different levels of dynamic range extension occur.
In Fig.~\ref{fig:dynamic range analysis}, we show this analysis using a simulated scene with $T_0 = 1/8000$s and $dT = 0.0005$s under a perfect permutation matrix.
We observe that the average dynamic range increases as the patch size increases, since larger patches are more likely to contain pixels with widely varying exposure times.}


\change{Patch-wise dynamic range analysis suggests that our method tends to struggle with scenes containing isolated highlight features. This is supported by the histogram plot, where 2×2 patches exhibit the lowest average dynamic range.
To further investigate this phenomenon, we propose a metric based on the Laplacian operator to quantify isolated highlight density:}

\change{
\begin{equation}
\label{eq:isolated highlight density}
\sum_{x,y} \max \left[-\nabla^2 \left(\frac{L(x,y)}{\sum_{x,y} L(x,y)}\right), 0 \right],
\end{equation}
}
\noindent\change{where $L$ is the luminance computed from the RGB image data. 
The isolated highlight density increases when more sharp intensity drops are present, as these are characteristic of isolated highlights in high dynamic range scenes.}

\begin{figure}
  \includegraphics[width=\linewidth]{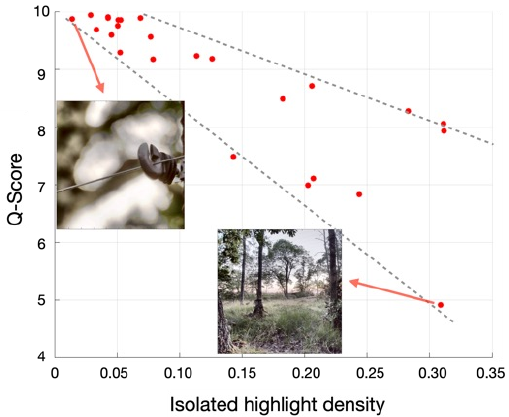}
  \caption{\change{Relationship between isolated highlight density and Q-score.
    Each red dot represents a scene from the dataset, plotted by its isolated highlight density (x-axis) and corresponding Q-score (y-axis). The isolated highlight density is computed using the Laplacian-based metric defined in Equation~\eqref{eq:isolated highlight density}, which captures spatially sharp, high-intensity transitions characteristic of isolated highlights in HDR scenes.
    Two representative scenes are annotated to illustrate the trend: scenes with spatially clustered highlights tend to achieve higher Q-scores, whereas scenes with sparse, isolated highlights created by sunlight filtering through tree branches show reduced reconstruction quality. This suggests our method is more effective when highlights are contiguous rather than sparse.}}
  \label{fig:sparsity_analysis}
\end{figure}

\change{We plot the computed isolated highlight density against Q-score in Fig.~\ref{fig:sparsity_analysis}. The results show a clear negative correlation, indicating that our method performs better when highlight regions are spatially clustered and struggles more with isolated highlight structures. Two representative scenes are annotated to illustrate this trend: one featuring clustered highlights caused by a defocused background, which results in a high Q-score; and another showing sparse, isolated highlights created by sunlight filtering through tree branches, which leads to reduced reconstruction quality.}
\subsection{Motion Blur}\label{sec:motion_blur_main_text}
\begin{figure}
\includegraphics{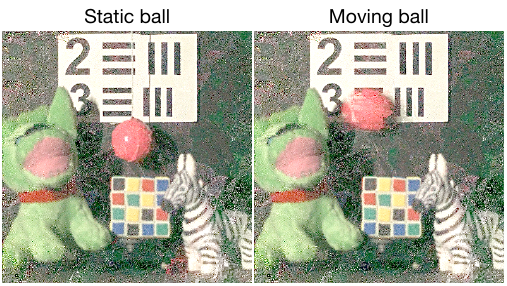}
    \caption{\change{Reconstructions comparing static (left) and dynamic (right) ball. The moving ball exhibits motion blur, but the rest of the scene is unaffected by the ball's motion.}}
    \label{fig:motion real data}
\end{figure}
\change{Our system is limited by the maximum exposure time that the GRR shutter function can provide.
As with other single-shot HDR methods, motion blur becomes a concern when longer exposures are required to capture low-intensity regions while preserving highlight details for inpainting.}

\change{To evaluate the impact of motion, we conducted a simulation (Appendix Fig.~\ref{fig:dynamic_simulation}) where the entire scene shifts by 1, 256, and 512 pixels within a 512×512 measurement.
The results show that our method can tolerate moderate motion, which manifests as mild noise in the reconstruction. However, when motion becomes severe, reconstruction quality deteriorates significantly.
We also provide an experimental comparison (Fig.~\ref{fig:motion real data}) using a scene with a static versus a moving ball in front of a fixed background. As expected, the moving ball appears blurred due to the fiber bundle being positioned at the image plane of the main lens, which causes the motion to be optically encoded in the measurement.}

\subsection{Depth sensitivity}\label{sec:depth_sensitivity}
\begin{figure}
\includegraphics{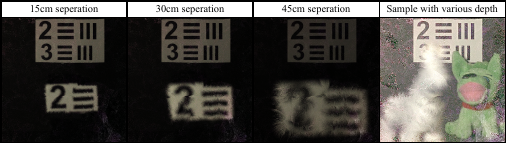}
    \caption{\change{Depth sensitivity of the proposed system under increasing defocus.
    Reconstructed images of two USAF resolution targets placed at different depths, with separations of 15cm, 30cm, and 45cm. 
    The rightmost panel shows a real-world scene with multiple objects at varying depths.}}
    \label{fig:optical blur real data}
\end{figure}
\change{Although our calibration procedure uses a 2D OLED display, our system's behavior over varying object distances is similar to that of a lens. This is because the fiber bundle input surface is positioned at the image plane of the main lens. Unlike methods that place a coding optic in the aperture plane, each fiber in the fiber bundle effectively integrates all light associated with a specific spatial location. That spatial location is shuffled, arriving at a pseudorandom pixel at the sensor. As a result, inverting the system matrix will simply recover an image with blur equivalent to that of the main lens image. This is in contrast to pupil-coded methods, wherein it is less clear what would be recovered if a single PSF were used to deblur a measurement taken from a different depth. In Appendix Fig.~\ref{fig:optical blur intro}, we show, using a light field-based visualization, that our system's response to an extended in-focus source is the same as its response to a defocused point source.}

\change{To validate this behavior, we conducted an experiment using two USAF resolution targets placed at different depths: one at the focal plane and the other gradually shifted away. As shown in Fig.~\ref{fig:optical blur real data}, the reconstructed image clearly captures both the focused and defocused targets, consistent with depth-dependent blurring in conventional lens systems.
We further evaluated the system using additional objects positioned at varying depths. These results confirm that, although defocused objects will be blurred in the reconstruction, in-focus areas are not affected--this is typical behavior for standard optical imaging, so we do not attempt to deblur out-of-focus areas.}
\subsection{Limitations}\label{sec:limitation}
\change{Our experiments utilize an unoptimized, low-cost, large-area fiber bundle to demonstrate the feasibility of our approach as a proof of concept.
The bundle contains over 800,000 multimode fibers with a numerical aperture (NA) of 0.66 and a diameter of 1.25 inches.
This large size introduces challenges in building a compact and efficient optical system, particularly with respect to alignment and light throughput, as illustrated in the prototype shown in Fig.~\ref{fig:prototype}.
We also note that the spatial mapping produced by our fiber bundle is significantly less random than an ideal permutation matrix. It exhibits visible clumping, where neighboring fibers tend to map to adjacent regions on the sensor.
In addition, the bundle contains a considerable number of broken fibers (approximately <5$\%$) and suffers from crosstalk between neighboring fibers. These imperfections—along with stray light in the relay optics—contribute to model mismatch and reduced calibration accuracy.
We find that accurate HDR reconstruction relies on precise knowledge of the system's transfer function, which is sensitive to typical HDR challenges such as sensor noise and optical stray light. These factors likely contribute to the degradation and artifacts observed in our reconstructions, particularly in very low-light regions of HDR scenes.}

\change{
As with any lens-based imaging system, our setup exhibits depth-dependent defocus blur in reconstruction, since the fiber bundle is positioned at the image plane.
Additional artifacts may result from mismatches between the fiber bundle’s NA and the 4-f relay optics, as well as from non-uniform transmittance across fibers, as shown in Fig.\ref{fig:optical blur real data}.
Similar to other single-shot HDR methods, our approach is not fully light-efficient: saturation and underexposure still occur, and inpainting is required.
Motion blur also poses a challenge, as shown in Fig.\ref{fig:motion real data}, becoming apparent when scene motion exceeds the maximum exposure time (i.e., the exposure time associated with the bottom row of the shutter function).
Finally, our explicit calibration procedure is time-consuming, requiring approximately 0.7 seconds per column of the permutation matrix $\mathbf{P}$.
This is primarily due to the non-injective nature of the system’s forward model, which arises from the combination of a large fiber bundle and the 4-f optical relay.}

\subsection{Future work}\label{sec:future work}
\change{Many of the artifacts and the time-consuming calibration process in our current system stem from imperfections in the hardware setup.
One promising direction for improvement is to carefully design the fiber arrangement using single-mode fibers to produce a pseudorandom output pattern.
Furthermore, the sensor could be directly bonded to the fiber bundle, eliminating the need for a bulky 4-f relay system.
This would result in a significantly more compact architecture with a more structured and predictable mapping between the object space and the sensor plane.
Such a configuration would move the system closer to an injective or near-permutation mapping, simplifying calibration by using binary patterns, potentially enabling instant calibration.
It also opens the door to deep learning-based blind reconstruction trained on diverse random mappings—an exciting direction for future exploration.
With improved system integration, this concept could be extended to applications such as endoscopy, where increasing the dynamic range of medical imaging is highly valuable.
Additionally, replacing the fiber bundle with exotic optical components such as diffractive optical elements or programmable masks could provide new pathways for compact, high-performance spatial multiplexing.
Together, these advancements can help evolve our method from a proof-of-concept prototype to a deployable, real-time, and application-specific HDR imaging system.}

\section{Conclusion}
In conclusion, we show that off-the-shelf sensors with GRR shutter can be combined with optical shuffling to achieve single-shot HDR imaging. We demonstrate this approach in simulation and in a real hardware prototype, which uses a random fiber bundle to shuffle the image. Compared to deep-learning methods, our approach shows approximately 25 dB of dynamic range extension using an 8-bit sensor. 
\section{Acknowledgment}
We acknowledge the use of the Lab Snacks® box in this work, which is a registered trademark of Thorlabs, Inc.
\bibliographystyle{ACM-Reference-Format}
\bibliography{bibliography}
\appendix
\section{Math derivative in spatial pseudorandomness in optical system}\label{section:appendixmath}
\subsection{Commute permutation matrix $\perm$ into nonlinear function $Q$}
The inverse of the random permutation matrix $\perm$ is equal to its transpose so that we can rewrite the first equation in Eq.~\eqref{eq:permcommuteq} as 
\begin{equation}\label{eq:commutedemo}
    \perm^T\meas=\perm^TQ\{\shutter\perm\object\}{.}
\end{equation}
We can express the input of $Q(\cdot)$ as a vector $\mathbf{y}$, where $\mathbf{y}=\shutter\perm\object$ and it can be written as a column vector with $n$ elements $\mathbf{y}=[y_1,y_2,\cdots,y_n]^\intercal$. 
The $\perm^T$ can be interpreted as an index mapping automorphism function $f(\cdot)$, which is bijective and has the same domain for the input and output.
As a result $$\perm^T \mathbf y = [y_{f(1)}, y_{f(2)},\ldots,y_{f(n)}]^\intercal{.}$$
In addition, because $Q(\cdot)$ is a pointwise operation, 
$$Q\{\mathbf y\} = [Q\{y_1\}, Q\{y_2\},\ldots,Q\{y_n\}]^\intercal{.}$$

Combining these properties, the right side of the Eq.~\eqref{eq:commutedemo} can be expressed as 
\begin{align*}
    \perm^TQ\{\mathbf{y}\}&=[Q(y_{f(1)}),Q(y_{f(2)}),\cdots,Q(y_{f(n)})]^\intercal\\
    &=Q\{\perm^T\mathbf{y}\}
\end{align*}

\subsection{Random exposure over the scene}
To validate that our method is equivalent to random exposure over the scene, we want to derive that the $\boldsymbol{\Lambda}$ is a diagonal matrix that commutes the diagonal value of the shutter matrix $\shutter$.
To begin with, we can express the diagonal shutter matrix $\shutter$ as $\shutter=\mbox{diag}\{s_{(1,1)}, s_{(2,2)},\cdots,s_{(m,m)} \}$ where the subscript means the index coordinate.
Right multiplying $\shutter$ by $\perm$ is equivalent to shuffling the column indices of $\shutter$ 
$$(\shutter \perm)_{(i,f(i))} = \shutter_{(i,i)}$$
\noindent where $f(i)$ is an automorphism function. 
We can also multiply the transpose of the same permutation matrix from the previous equation to map the row index to another index value which can be expressed as $\perm^T\shutter\perm=\mbox{diag}\{s_{(f(1),f(1))}, s_{(f(2),f(2))},\cdots,s_{(f(m),f(m))} \}$. 
Because the automorphism function is bijective and has the same domain for the input and output, the $\boldsymbol{\Lambda}=\perm^T\shutter\perm$ is a diagonal matrix which permutes the diagonal value of the shutter function $\shutter$.
\section{Simulation results}
\subsection{Ablation test results}
Table.~\ref{tab:ablation test} has detailed information on the ablation test using various optical, shutter functions, and prior settings with different metrics.
\subsection{Baseline test results}
Table.~\ref{tab:baseline test} has detailed information on the baseline test using HDR CNN, Deep Optics HDR, spatially varying exposure, and our method with different metrics. And the reconstruction image and ground truth are normalized by their maximum values.

\begin{figure}
    \centering
    \includegraphics{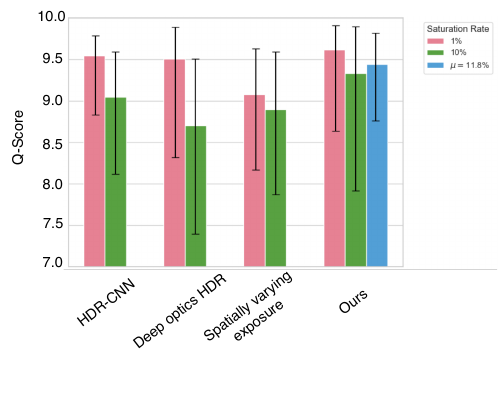}
    \caption{\change{Baseline test comparison Q-score. Bar plot of Q-score reconstruction as Fig.~\ref {fig:baselinetestplot}, but the ground truth and reconstruction are normalized by their mean values before sending to metrics. }}
    \label{fig:baseline test norm by mean}
\end{figure}

\change{Table.~\ref{tab:baseline test mean norm} has the same methods with the same metrics as Table.~\ref{tab:baseline test}, but the ground truth and reconstruction image are normalized to their mean values. 
A bar plot Fig.~\ref{fig:baseline test norm by mean} shows similar results for better visualization. }

\begin{table*}
  \caption{\change{Ablation tests using different forward models}}
  \label{tab:ablation test}
  \centering
   \resizebox{\textwidth}{!}{
  \begin{tabular}{c|M{23mm}|c|c|c|c|c|c|c|c|c|c}
    \toprule
    \textbf{Method} & \textbf{Saturation Rate} & \textbf{Prior} & \textbf{Stat} 
    & \textbf{PSNR}($\mathbf{\uparrow}$)(dB) & \multicolumn{1}{c|}{\textbf{PSNR$_\gamma$}($\mathbf{\uparrow}$)(dB)} 
    & \textbf{SSIM} ($\mathbf{\uparrow}$)& \multicolumn{1}{c|}{\textbf{MSSSIM}($\mathbf{\uparrow}$)} 
    & \textbf{VSI} ($\mathbf{\uparrow}$)& \multicolumn{1}{c|}{\textbf{FSIM}($\mathbf{\uparrow}$)} 
    & \textbf{Q} ($\mathbf{\uparrow}$)& \multicolumn{1}{c}{\textbf{$\mathbf{Q_{JOD}}$}($\mathbf{\uparrow}$)} \\
    \cline{1-12}

    \multirow{12}{*}{Global shutter with lens} 
    & \multirow{2}{*}{3\%} & \multirow{2}{*}{No} & Mean 
    & 12.96 & 15.02 & 0.72 & 0.81 & 0.93 & 0.86 & 6.32 & 7.21 \\
    \cline{4-12}
    &                        &                      & Std  
    & 2.23  & 5.56  & 0.23 & 0.16 & 0.05 & 0.10 & 0.91 & 0.86 \\
    \cline{2-12}
    
    & \multirow{2}{*}{5\%} & \multirow{2}{*}{No} & Mean 
    & 11.42 & 13.13 & 0.65 & 0.74 & 0.91 & 0.82 & 5.57 & 6.47 \\
    \cline{4-12}
    &                        &                      & Std  
    & 1.87  & 4.72  & 0.25 & 0.18 & 0.05 & 0.11 & 0.99 & 0.98 \\
    \cline{2-12}

    & \multirow{2}{*}{10\%} & \multirow{2}{*}{No} & Mean 
    & 9.50 & 10.62 & 0.52 & 0.58 & 0.89 & 0.77 & 4.60 & 5.47 \\
    \cline{4-12}
    &                         &                      & Std  
    & 1.41 & 3.64 & 0.26 & 0.24 & 0.05 & 0.12 & 0.88 & 0.94 \\
    \cline{2-12}

    & \multirow{2}{*}{3\%} & \multirow{2}{*}{Yes} & Mean 
    & 23.94 & 24.09 & 0.89 & 0.94 & 0.98 & 0.95 & 9.03 & 9.40 \\
    \cline{4-12}
    &                        &                      & Std  
    & 8.23  & 8.80  & 0.16 & 0.19 & 0.04 & 0.08 & 1.07 & 0.98 \\
    \cline{2-12}

    & \multirow{2}{*}{5\%} & \multirow{2}{*}{Yes} & Mean 
    & 21.54 & 21.58 & 0.85 & 0.92 & 0.97 & 0.93 & 8.63 & 9.11 \\
    \cline{4-12}
    &                        &                      & Std  
    & 7.37  & 7.65  & 0.19 & 0.12 & 0.04 & 0.09 & 1.24 & 0.96 \\
    \cline{2-12}

    & \multirow{2}{*}{10\%} & \multirow{2}{*}{Yes} & Mean 
    & 18.11 & 18.10 & 0.78 & 0.87 & 0.95 & 0.90 & 7.63 & 8.32 \\
    \cline{4-12}
    &                         &                      & Std  
    & 6.02  & 6.29  & 0.23 & 0.15 & 0.05 & 0.11 & 1.41 & 1.19 \\
    \cline{1-12}

    \multirow{16}{*}{GRR shutter with lens} 
    & \multirow{2}{*}{3\%} & \multirow{2}{*}{No} & Mean 
    & 26.73 & 26.95 & 0.93 & 0.95 & 0.98 & 0.96 & 5.55 & 6.39 \\
    \cline{4-12}
    &                        &                      & Std  
    & 6.15  & 6.46  & 0.12 & 0.07 & 0.03 & 0.05 & 1.60 & 1.60 \\
    \cline{2-12}

    & \multirow{2}{*}{5\%} & \multirow{2}{*}{No} & Mean 
    & 25.41 & 25.66 & 0.91 & 0.94 & 0.98 & 0.95 & 5.50 & 6.35 \\
    \cline{4-12}
    &                        &                      & Std  
    & 5.57  & 5.86  & 0.13 & 0.07 & 0.03 & 0.05 & 1.56 & 1.57 \\
    \cline{2-12}

    & \multirow{2}{*}{10\%} & \multirow{2}{*}{No} & Mean 
    & 22.94 & 23.16 & 0.88 & 0.92 & 0.97 & 0.93 & 5.29 & 6.16 \\
    \cline{4-12}
    &                         &                      & Std  
    & 4.65  & 4.91  & 0.14 & 0.07 & 0.03 & 0.06 & 1.40 & 1.44 \\
    \cline{2-12}

    & \multirow{2}{*}{3\%} & \multirow{2}{*}{Yes} & Mean 
    & 28.92 & 29.54 & 0.94 & 0.96 & 0.98 & 0.97 & 7.54 & 8.30 \\
    \cline{4-12}
    &                        &                      & Std  
    & 6.63  & 7.12  & 0.11 & 0.07 & 0.03 & 0.05 & 0.99 & 0.87 \\
    \cline{2-12}

    & \multirow{2}{*}{5\%} & \multirow{2}{*}{Yes} & Mean 
    & 27.24 & 27.70 & 0.92 & 0.95 & 0.98 & 0.96 & 7.46 & 8.24 \\
    \cline{4-12}
    &                        &                      & Std  
    & 6.33  & 6.74  & 0.14 & 0.08 & 0.03 & 0.06 & 0.95 & 0.83 \\
    \cline{2-12}

    & \multirow{2}{*}{10\%} & \multirow{2}{*}{Yes} & Mean 
    & 23.74 & 23.94 & 0.88 & 0.92 & 0.97 & 0.93 & 7.04 & 7.88 \\
    \cline{4-12}
    &                         &                      & Std  
    & 5.58  & 5.79  & 0.16 & 0.09 & 0.04 & 0.07 & 0.93 & 0.85 \\
        \cline{2-12}

    & \multirow{2}{*}{$\mu=8.8\%$} & \multirow{2}{*}{No} & Mean 
    & 31.28 & 31.81 & 0.96 & 0.96 & \textbf{0.99} & 0.96 & 5.77 & 6.64 \\
    \cline{4-12}
    &                         &                      & Std  
    & 8.44  & 8.79  & 0.04 & 0.05 & 0.01 & 0.04 & 1.28 & 1.25 \\
        \cline{2-12}

    & \multirow{2}{*}{$\mu=8.8\%$} & \multirow{2}{*}{Yes} & Mean 
    & 31.20 & 32.24 & 0.96 & 0.96 & \textbf{0.99} & 0.96 & 6.75 & 7.58 \\
    \cline{4-12}
    &                         &                      & Std  
    & 7.90  & 8.57  & 0.04 & 0.05 & 0.01 & 0.03 & 1.23 & 1.15 \\
    \cline{1-12}

    \multirow{8}{*}{\textbf{Ours}} 
    & \multirow{2}{*}{3\%} & \multirow{2}{*}{Yes} & Mean 
    & \textbf{38.28} & 39.66 & \textbf{0.99} & \textbf{0.99} & \textbf{0.99} & \textbf{0.99} & \textbf{9.56} & \textbf{9.80} \\
    \cline{4-12}
    &                        &                      & Std  
    & 5.51  & 6.35  & 0.01 & 0.01 & 0.01 & 0.01 & 0.41 & 0.26 \\
    \cline{2-12}

    & \multirow{2}{*}{5\%} & \multirow{2}{*}{Yes} & Mean 
    & 36.86 & 38.05 & \textbf{0.99} & \textbf{0.99} & \textbf{0.99} & \textbf{0.99} & 9.49 & 9.75 \\
    \cline{4-12}
    &                        &                      & Std  
    & 5.72  & 6.55  & 0.02 & 0.01 & 0.02 & 0.01 & 0.51 & 0.34 \\
    \cline{2-12}

    & \multirow{2}{*}{10\%} & \multirow{2}{*}{Yes} & Mean 
    & 38.12 & \textbf{39.83} & 0.98 & \textbf{0.99} & \textbf{0.99} & \textbf{0.99} & 9.33 & 9.65 \\
    \cline{4-12}
    &                        &                      & Std  
    & 5.66  & 6.61  & 0.02 & 0.01 & 0.02 & 0.02 & 0.62 & 0.42 \\
    \cline{2-12}

    & \multirow{2}{*}{$\mu=11.8\%$} & \multirow{2}{*}{Yes} & Mean 
    & 37.79 & 38.82 & \textbf{0.99} & \textbf{0.99} & \textbf{0.99} & \textbf{0.99} & 9.43 & 9.73 \\
    \cline{4-12}
    &                                  &                    & Std  
    & 6.18  & 6.70  & 0.02 & 0.01 & 0.01 & 0.01 & 0.37 & 0.23 \\
    \bottomrule
  \end{tabular}
  }
\end{table*}

\begin{table*}
  \caption{\change{Baseline tests using different single-shot HDR methods normalized by maximum}}
  \label{tab:baseline test}
  \centering
  \resizebox{\textwidth}{!}{
  \begin{tabular}{c|M{23mm}|c|c|c|c|c|c|c|c|c}
    \toprule
    \textbf{Method} & \textbf{Saturation Rate} & \textbf{Stat} 
    & \textbf{PSNR}($\mathbf{\uparrow}$)(dB) & \multicolumn{1}{c|}{\textbf{PSNR$_\gamma$}($\mathbf{\uparrow}$)(dB)} 
    & \textbf{SSIM}($\mathbf{\uparrow}$) & \multicolumn{1}{c|}{\textbf{MSSSIM}($\mathbf{\uparrow}$)} 
    & \textbf{VSI}($\mathbf{\uparrow}$) & \multicolumn{1}{c|}{\textbf{FSIM}($\mathbf{\uparrow}$)} 
    & \textbf{Q}($\mathbf{\uparrow}$) & \multicolumn{1}{c}{\textbf{$\mathbf{Q_{JOD}}$}($\mathbf{\uparrow}$)} \\
    \cline{1-11}

    \multirow{4}{*}{Spatially varying exposure} 
    & \multirow{2}{*}{1\%} &  Mean 
    & 37.59 & 39.14 & 0.98 & \textbf{0.99} & \textbf{0.99} & \textbf{0.99} & 9.06 & 9.50 \\
    \cline{3-11}
    &                        &  Std  
    & 4.62  & 5.04  & 0.02 & 0.01 & 0.01 & 0.01 & 0.44 & 0.30 \\
    \cline{2-11}

    & \multirow{2}{*}{10\%} &  Mean 
    & 33.59 & 34.98 & 0.97 & \textbf{0.99} & \textbf{0.99} & \textbf{0.99} & 8.87 & 9.36 \\
    \cline{3-11}
    &                        &   Std  
    & 4.63  & 5.20  & 0.02 & 0.01 & 0.01 & 0.01 & 0.64 & 0.48 \\
    \cline{1-11}

    \multirow{4}{*}{HDR CNN} 
    & \multirow{2}{*}{1\%} &  Mean 
    & 25.25 & 25.29 & 0.90 & 0.94 & 0.98 & 0.95 & 9.15 & 9.53 \\
    \cline{3-11}
    &                        &    Std  
    & 7.85  & 7.97  & 0.16 & 0.09 & 0.03 & 0.07 & 0.70 & 0.50 \\
    \cline{2-11}

    & \multirow{2}{*}{10\%} &  Mean 
    & 22.16 & 22.37 & 0.86 & 0.92 & 0.97 & 0.93 & 8.75 & 9.28 \\
    \cline{3-11}
    &                        &     Std  
    & 6.73  & 7.14  & 0.16 & 0.09 & 0.03 & 0.07 & 0.65 & 0.48 \\
    \cline{1-11}

    \multirow{4}{*}{Deep Optics HDR} 
    & \multirow{2}{*}{1\%} &  Mean 
    & 25.94 & 26.01 & 0.88 & 0.93 & 0.97 & 0.95 & 9.06 & 9.45 \\
    \cline{3-11}
    &                        & Std  
    & 8.74  & 8.95  & 0.18 & 0.11 & 0.04 & 0.09 & 0.89 & 0.64 \\
    \cline{2-11}

    & \multirow{2}{*}{10\%} &  Mean 
    & 19.18 & 19.30 & 0.72 & 0.83 & 0.93 & 0.86 & 8.14 & 8.78 \\
    \cline{3-11}
    &                        &Std  
    & 7.72  & 8.08  & 0.27 & 0.18 & 0.07 & 0.15 & 1.02 & 0.83 \\
    \cline{1-11}
    \multirow{6}{*}{\textbf{Ours}} 
    & \multirow{2}{*}{1\%} &  Mean 
    & \textbf{40.32} & \textbf{42.07} & \textbf{0.99} & \textbf{0.99} & \textbf{0.99} & \textbf{0.99} & \textbf{9.60} & \textbf{9.81} \\
    \cline{3-11}
    &                        &   Std  
    & 5.29  & 6.17  & 0.01 & 0.01 & 0.01 & 0.01 & 0.41 & 0.25 \\
    \cline{2-11}

    & \multirow{2}{*}{10\%} &  Mean 
    & 35.09 & 36.02 & 0.98 & \textbf{0.99} & \textbf{0.99} & \textbf{0.99} & 9.30 & 9.62 \\
    \cline{3-11}
    &                        &     Std  
    & 5.81  & 6.52  & 0.02 & 0.02 & 0.02 & 0.02 & 0.69 & 0.48 \\
    \cline{2-11}

    & \multirow{2}{*}{$\mu=11.8\%$} & Mean 
    & 37.79 & 38.82 & \textbf{0.99} & \textbf{0.99} & \textbf{0.99} & \textbf{0.99} & 9.43 & 9.73 \\
    \cline{3-11}
    &                                  &  Std  
    & 6.18  & 6.70  & 0.02 & 0.01 & 0.01 & 0.01 & 0.37 & 0.23 
    \\
    \bottomrule
  \end{tabular}
  }
\end{table*}

\begin{table*}
  \caption{\change{Baseline tests using different single-shot HDR methods normalized by mean}}
  \label{tab:baseline test mean norm}
  \centering
  \resizebox{\textwidth}{!}{
  \begin{tabular}{c|M{23mm}|c|c|c|c|c|c|c|c|c}
    \toprule
    \textbf{Method} & \textbf{Saturation Rate} & \textbf{Stat} 
    & \textbf{PSNR}($\mathbf{\uparrow}$)(dB) & \multicolumn{1}{c|}{\textbf{PSNR$_\gamma$}($\mathbf{\uparrow}$)(dB)} 
    & \textbf{SSIM}($\mathbf{\uparrow}$) & \multicolumn{1}{c|}{\textbf{MSSSIM}($\mathbf{\uparrow}$)} 
    & \textbf{VSI}($\mathbf{\uparrow}$) & \multicolumn{1}{c|}{\textbf{FSIM}($\mathbf{\uparrow}$)} 
    & \textbf{Q}($\mathbf{\uparrow}$) & \multicolumn{1}{c}{\textbf{$\mathbf{Q_{JOD}}$}($\mathbf{\uparrow}$)} \\
    \cline{1-11}

    \multirow{4}{*}{Spatially varying exposure} 
    & \multirow{2}{*}{1\%} &  Mean 
    & 39.27 & 41.46 & 0.98 & \textbf{0.99} & \textbf{0.99} & \textbf{0.99} & 9.08 & 9.51 \\
    \cline{3-11}
    &                        &  Std  
    & 4.29  & 4.65  & 0.01 & 0.01 & 0.01 & 0.01 & 0.44 & 0.30 \\
    \cline{2-11}

    & \multirow{2}{*}{10\%} &  Mean 
    & 35.95 & 37.95 & 0.97 & \textbf{0.99} & \textbf{0.99} & \textbf{0.99} & 8.90 & 9.39 \\
    \cline{3-11}
    &                        &   Std  
    & 4.14  & 4.67  & 0.02 & 0.01 & 0.01 & 0.01 & 0.61 & 0.46 \\
    \cline{1-11}

    \multirow{4}{*}{HDR CNN} 
    & \multirow{2}{*}{1\%} &  Mean 
    & 37.89 & 38.35 & \textbf{0.99} & \textbf{0.99} & \textbf{0.99} & \textbf{0.99} & 9.55 & 9.80 \\
    \cline{3-11}
    &                        &    Std  
    & 2.90  & 3.07  & 0.01 & 0.01 & 0.01 & 0.01 & 0.31 & 0.19 \\
    \cline{2-11}

    & \multirow{2}{*}{10\%} &  Mean 
    & 32.87 & 33.69 & 0.98 & 0.98 & 0.98 & 0.98 & 9.05 & 9.50 \\
    \cline{3-11}
    &                        &     Std  
    & 3.13  & 3.56  & 0.01 & 0.01 & 0.02 & 0.02 & 0.46 & 0.32 \\
    \cline{1-11}

    \multirow{4}{*}{Deep Optics HDR} 
    & \multirow{2}{*}{1\%} &  Mean 
    & \textbf{43.76} & 45.02 & \textbf{0.99} & \textbf{0.99} & \textbf{0.99} & \textbf{0.99} & 9.51 & 9.76 \\
    \cline{3-11}
    &                        & Std  
    & 3.67  & 4.38  & 0.01 & 0.01 & 0.01 & 0.01 & 0.52 & 0.34 \\
    \cline{2-11}

    & \multirow{2}{*}{10\%} &  Mean 
    & 34.14 & 34.98 & 0.98 & 0.98 & 0.97 & 0.97 & 8.70 & 9.24 \\
    \cline{3-11}
    &                        &Std  
    & 4.31  & 5.14  & 0.01 & 0.01 & 0.04 & 0.02 & 0.71 & 0.55 \\
    \cline{1-11}

    \multirow{6}{*}{\textbf{Ours}} 
    & \multirow{2}{*}{1\%} &  Mean 
    & 43.43 & \textbf{46.60} & \textbf{0.99} & \textbf{0.99} & \textbf{0.99} & \textbf{0.99} & \textbf{9.62} & \textbf{9.83} \\
    \cline{3-11}
    &                        &   Std  
    & 4.30  & 4.85  & 0.01 & 0.01 & 0.01 & 0.01 & 0.38 & 0.24 \\
    \cline{2-11}

    & \multirow{2}{*}{10\%} &  Mean 
    & 38.12 & 39.83 & 0.98 & \textbf{0.99} & \textbf{0.99} & \textbf{0.99} & 9.33 & 9.65 \\
    \cline{3-11}
    &                        &     Std  
    & 5.66  & 6.61  & 0.02 & 0.01 & 0.02 & 0.02 & 0.62 & 0.42 \\
    \cline{2-11}

    & \multirow{2}{*}{$\mu=11.8\%$} & Mean 
    & 40.61 & 42.39 & \textbf{0.99} & \textbf{0.99} & \textbf{0.99} & \textbf{0.99} & 9.44 & 9.73 \\
    \cline{3-11}
    &                                  &  Std  
    & 5.52  & 5.93  & 0.01 & 0.01 & 0.01 & 0.01 & 0.36 & 0.23 \\
    \bottomrule
  \end{tabular}
  }
\end{table*}

\section{Experimental results}
\subsection{Experimental setup}
\begin{figure}
    \centering
    \includegraphics{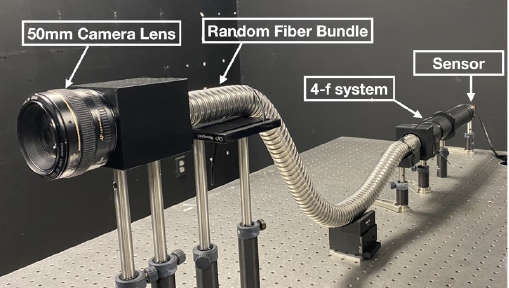}
    \caption{\change{The photo of the prototype. The image system starts with a 50 mm camera lens, followed by a random fiber bundle and a 4-f relay optical system to generate an image on the Allied Vision 1800 U-1240 (Sony IMX226).}}
    \label{fig:prototype}
\end{figure}

\change{Fig.~\ref{fig:prototype}, shows the prototype of our system.}

\subsection{Different orientation tests for quantitative validation}
\begin{figure}
    \centering
    \includegraphics{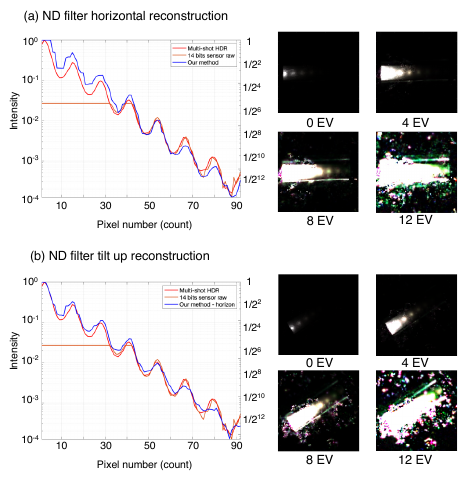}
    \caption{The reconstruction of ND filters with two orientations. (a) The intensity analysis of the horizontal ND filter and reconstruction with various EVs. (b) The intensity analysis of the tilt-up ND filter and reconstruction with various EVs.}
    \label{fig:two_orientation_ndfilter}
\end{figure}
To validate whether the sample's orientation will impact our method's reconstruction dynamic range. 
We test two different orientations of the ND filter sample and display the intensity analysis in Fig.\ref{fig:two_orientation_ndfilter}.
\begin{figure}
\includegraphics{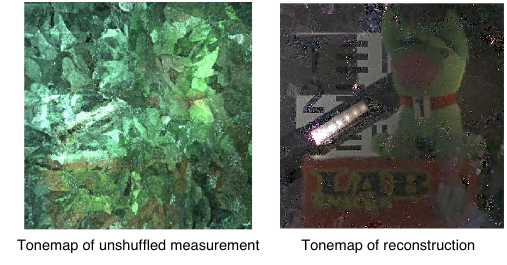}
    \caption{The comparison is between the unshuffled measurement and reconstruction from the real system. On the left side is the tonemap image of the unshuffled measurement ($\perm^\intercal \meas$). On the right side is the tonemap of the reconstruction image of the same sample. The sample Lab Snacks\textsuperscript{\textregistered} box used in reconstruction is a registered trademark of Thorlabs, Inc.}
    \label{fig:realsampleshuffle}
\end{figure}

\subsection{Unshuffle experimental measurement}
In the real system, we can unshuffle the measurement to obtain the random exposure map of the object. 
We can still see a rough structure in Fig.~\ref{fig:realsampleshuffle} of the scene, but the quality is far from that of the reconstruction.
Because the noise term will be contained, unlike the ideal situation, and the optical matrix $\perm$ is not a random permutation matrix, we cannot insert the unshuffle matrix $\perm^T$ inside of the clip function. 

\begin{figure*}
\includegraphics{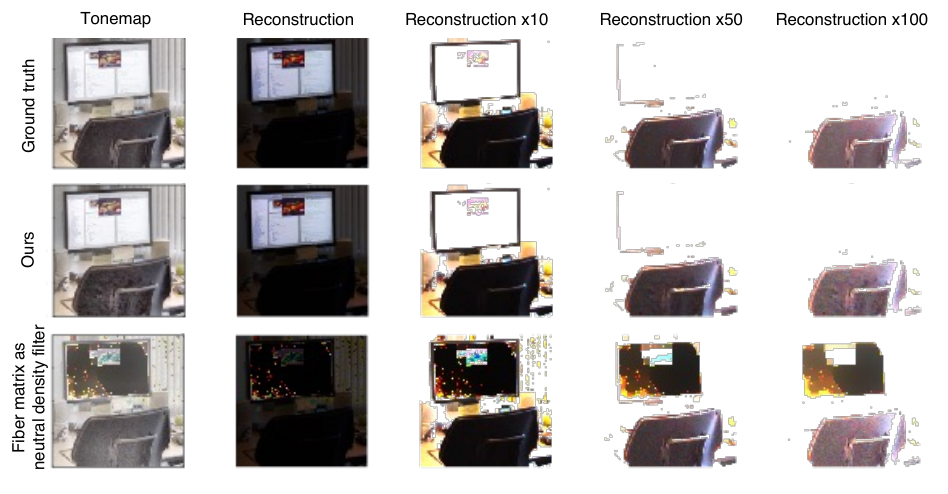}
    \caption{Simulation test images when we regard the optical matrix of the random fiber bundle as a neutral density filter. We display the tonemap, reconstruction, and reconstruction with different scales of our method and the fiber matrix as a neutral density filter compared with the ground truth image.}
    \label{fig:fiberproperty test}
\end{figure*}
\subsection{Real optical matrix of random fiber bundle simulation}
We include a simulation test in Fig.\ref{fig:fiberproperty test} regarding the optical matrix as a neutral density filter to show that the energy attenuation by the random fiber bundle does not impact the HDR extension. 
\section{Optical Blur}
\begin{figure}
\includegraphics{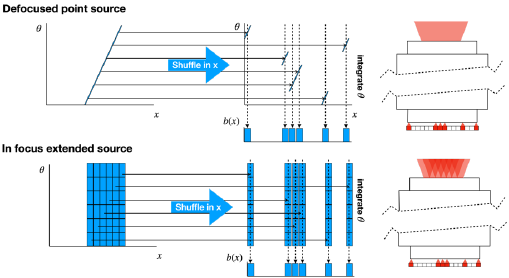}
    \caption{\change{Conceptual illustration of the effect of optical blur in the proposed system.
    (a) A defocused point source produces a spatially distributed pattern across the sensor due to optical blur.
    (b) An in-focus set of extended point sources can reproduce the same measurement pattern, supporting the idea that in-focus calibration can represent the response to defocused input.}}
    \label{fig:optical blur intro}
\end{figure}

\change{Fig.~\ref{fig:optical blur intro} presents a conceptual illustration showing that the effect of optical blur can be approximated by an all-in-focus calibration. Here, we show that the systems' response to an extended in-focus source is equivalent to its response from a defocused point source. We show a light field epipolar sketch of each source. The fiber bundle permutes the spatial dimension, then the sensor integrates the angular dimension. As a result, all that matters is the spatial distribution of the light field arriving at the fiber bundle; its angular content does not matter. Therefore, an in-focus extended source with spatial intensity equal to that of a defocused point source produces identical measurements. This means we would expect objects captured from outside our calibration plane to be reconstructed with a blur similar to that of the primary lens. Note, for out-of-focus objects, we are satisfied with recovering blurred results and do not attempt to deblur the image. Note that our 4-f system has an aperture that is smaller than that of the fibers in the bundle. As a result, some high-angle light exiting the fiber may be clipped, causing our system to deviate slightly from this prediction. In a future system, where the bundle is bonded directly to a sensor, this would not be an issue.}

\section{Motion blur}
\begin{figure}
\includegraphics{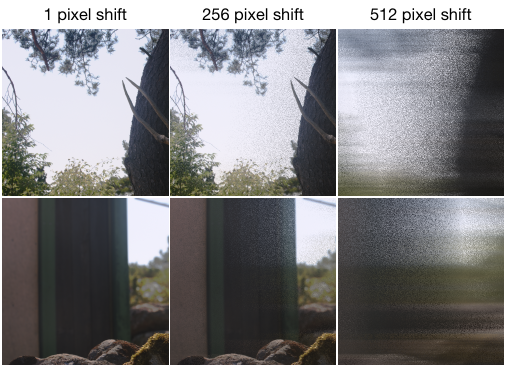}
    \caption{\change{Simulation of reconstruction under varying scene motion.
    Each column shows reconstruction results when a global pixel shift of 1, 256, and 512 pixels is applied between two samples.}}
    \label{fig:dynamic_simulation}
\end{figure}

\change{Fig.~\ref{fig:dynamic_simulation} shows simulation results where the entire scene is shifted by 1, 256, and 512 pixels across two 512×512 measurements to evaluate the impact of motion on reconstruction quality.}

\section{Comparison with Spatially Varying Exposure}
\begin{figure}
\includegraphics{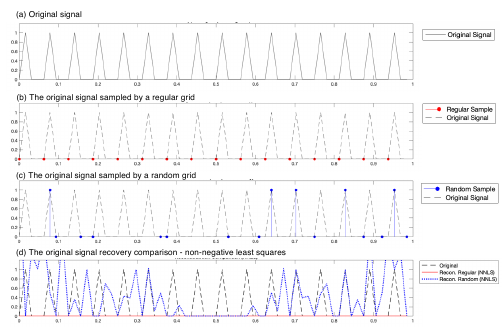}
    \caption{\change{ 1D signal sampling and reconstruction comparison between regular and random grids.
    (a) The original high-frequency signal.
    (b) The signal sampled 25$\%$ uniformly on a regular grid, leading to aliasing.
    (c) The same signal sampled 25$\%$ using a random grid, which disrupts aliasing patterns.
    (d) Reconstruction results using non-negative least squares (NNLS). The regular grid produces strong aliasing and poor recovery (red), while the random sampling better preserves high-frequency content and results in a more accurate reconstruction (blue), demonstrating the advantage of incoherent sampling in the context of compressive sensing.}}
    \label{fig:fre_analysis_1d}
\end{figure}

\change{Appendix Fig.~\ref{fig:fre_analysis_1d} illustrates a 1D signal sampled at 25$\%$ density using both regular and random grids, and compares the corresponding reconstruction results.
This example highlights the aliasing artifacts caused by regular sampling and demonstrates how random sampling, when combined with a simple non-negative least squares (NNLS) reconstruction, better preserves high-frequency content—supporting the effectiveness of incoherent sampling strategies in our system design.}
\section{Dynamic Range Dependency to the $\mathbf{T_0}$ and $\mathbf{T_r}$}
\begin{figure}
\includegraphics{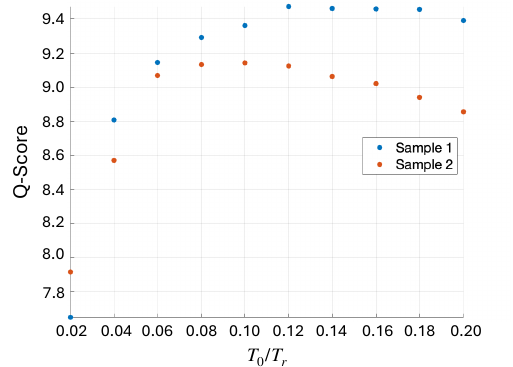}
    \caption{\change{The plot shows the Q-score of two simulated samples using permutation matrix(Sample 1 and Sample 2) as a function of the ratio between the baseline exposure time ($T_0$) and the readout time per row ($t_r$).}}
    \label{fig:HDR_t0_tr}
\end{figure}

Appendix Fig. \ref{fig:HDR_t0_tr} illustrates the Q-score of two simulated samples using a permutation matrix as a function of the ratio between the baseline exposure time ($T_0$) and the readout time per row ($t_r$).
\section{Challenges in Generating HDR Ground Truth for Training}\label{sec:HDR ground truth}
\change{Instead of displaying structured patterns, we opted to use a white dot as the calibration target due to the inherent limitations of modern HDR displays. 
Most HDR displays either suffer from elevated black levels (as in LCDs) or apply undisclosed gray-level modifications such as automatic brightness limiting or tone mapping, which are designed for perceptual quality rather than photometric accuracy.
These limitations make it challenging to generate reliable HDR ground truth with units of radiant flux, which is essential for supervised deep learning. }
\end{document}